\documentclass[sigconf,nonacm]{acmart}
\AtBeginDocument{%
  }

\def\code#1{\texttt{#1}}

\setcopyright{none}
\usepackage{amsmath}
\usepackage{multirow}
\usepackage{booktabs}
\usepackage{comment}
\usepackage{graphicx}
\usepackage{subcaption}  
\usepackage{float}
\usepackage{xcolor}
\usepackage{enumitem}
\usepackage{url}
\usepackage{siunitx}
\usepackage{tikz}
\usetikzlibrary{positioning}
\usetikzlibrary{shapes.misc}

\DeclareMathOperator*{\argmin}{arg\,min}

\newcommand{\circlednum}[1]{%
  \tikz[baseline=(char.base)]{
    \node[shape=circle,fill=black,inner sep=1pt] (char) {\color{white}\small #1};
  }%
}

\begin{document}

\title{DEFT: Joint Task Placement and DVFS for Energy-Efficient Multi-GPU Runtimes}

\author{Jing Chen}
\orcid{0000-0003-3409-8651}
\affiliation{%
  \institution{Chalmers University of Technology and University of Gothenburg}
  \city{Gothenburg}
  \country{Sweden}}
\email{chjing@chalmers.se}

\author{Miquel Pericàs}
\orcid{0000-0002-7583-6609}
\affiliation{%
  \institution{Chalmers University of Technology and University of Gothenburg}
  \city{Gothenburg}
  \country{Sweden}}
\email{miquelp@chalmers.se}



\begin{abstract}
% logical flow of the Introduction: Problem (Energy/DVFS overheads) => Gap (Lack of joint optimization/granularity awareness) => Solution (CUDASTF + Cost Model) => Methodology (Profitability analysis) => Results.

% Energy efficiency has become a first-order design constraint in modern high-performance computing systems as large-scale applications increasingly rely on multi-GPU nodes to deliver high throughput.
Energy efficiency has become a first-order concern in modern high-performance computing systems, as it directly determines achievable throughput under fixed power budgets. 
% as improving energy efficiency directly increases achievable throughput per watt in multi-GPU workloads.
% Although Dynamic Voltage and Frequency Scaling (DVFS) provides an effective mechanism for reducing GPU energy consumption, existing runtime systems typically decouple DVFS from task placement and inter-GPU communication, focus on single-GPU execution, or lack principled mechanisms to adapt frequency decisions to task granularity and runtime contention in multi-GPU environments.
Although Dynamic Voltage and Frequency Scaling (DVFS) provides an effective mechanism for reducing GPU energy consumption, existing runtime systems decouple DVFS from task placement and inter-GPU communication, focus on single-GPU execution, or cannot adapt frequency to task granularity and runtime contention in multi-GPU environments.
Consequently, current schedulers fail to capture the tight coupling between task placement, frequency selection, and inter-GPU data movement that fundamentally governs energy–performance trade-offs on multi-GPU systems.

This paper presents DEFT, an energy-aware scheduling framework that jointly optimizes task-to-device assignment and per-GPU DVFS configuration for task-based multi-GPU applications.
DEFT employs a cost-model–driven strategy that integrates slack awareness, throughput awareness, and explicit modeling of task execution cost, inter-GPU data movement, and DVFS transition overheads, enabling coordinated placement and frequency decisions at task granularity under dynamic runtime conditions.
% DEFT employs a cost-model–driven strategy integrating slack awareness, throughput awareness, and explicit modeling of task execution cost, inter-GPU data movement, and DVFS transition overheads, coordinating placement and frequency at task granularity under dynamic runtime conditions.
We prototype DEFT within the CUDASTF runtime and demonstrate its effectiveness across five optimization objectives.
% and evaluate it on NVIDIA L4 and L40S multi-GPU systems.
% DEFT integrates slack awareness, throughput awareness, and explicit modeling of inter-GPU data transfer and DVFS transition overheads within a unified cost model, enabling coordinated placement and frequency decisions that adapt to both task granularity and dynamic system conditions.
% We implement DEFT in the CUDASTF runtime and evaluate it on NVIDIA L4 and L40S multi-GPU systems across various energy performance trade-off metrics.
The evaluation shows that DEFT reduces energy consumption by 14.8\% and 4.8\% on average on NVIDIA L40S and L4, and reduces EDP by 9.9\% and 3.7\%, respectively, while maintaining performance within 1.5\% of the fastest baseline.
\end{abstract}




\maketitle

\section{Introduction} \label{sec:intro}

\begin{comment}
The structure of the Introduction: 
1. Problem statement (energy efficiency in HPC, multi-GPU) 
2. DVFS opportunities and challenges 
3. Multi-GPU complexity 
4. Task-based models and CUDASTF 
5. Our approach (DEFT description) 
6. Contributions 
\end{comment}

%%%%% Energy efficiency, GPU %%%%%
As High Performance Computing (HPC) systems advance beyond the Exascale barrier, energy efficiency has emerged as a primary constraint limiting further scalability~\cite{Shehabi2024DataCenterEnergy, TCO_HPC_Report, DataCenterEnergy2020, DVFSGovernor2015}.
% Under fixed power budgets, as is typical in data centers, energy efficiency directly determines achievable throughput: more efficient execution enables greater computation per watt within the same power envelope.
% In large-scale deployments, improving energy efficiency is critical for sustaining throughput growth, as it directly translates to greater computation delivered per watt for a given power budget.
In large-scale deployments, improved energy efficiency directly translates into higher aggregate throughput. 
% improved energy efficiency directly translates into higher throughput per watt.
% Contemporary HPC nodes increasingly rely on dense GPU integration, with multiple high-performance GPUs deployed within a single node to provide massive computational throughput.
% Contemporary HPC nodes increasingly rely on multi-GPU integration to provide massive computational throughput.
% However, translating this raw computational capability into energy-efficient execution remains challenging.
Contemporary HPC nodes increasingly rely on dense multi-GPU integration to deliver massive computational capability, yet translating this capability into energy-efficient execution remains challenging.
In practice, GPUs are often overprovisioned relative to application-level parallelism~\cite{GPUUtilPEARC2025}, and default power management policies tend to maintain peak operating frequencies whenever devices are allocated, even when utilization is low~\cite{GPUPowerManagement2023}, resulting in substantial energy waste.
%  or cores are temporarily idle

% Furthermore, energy optimization differs fundamentally from performance optimization in task-based systems.
% While energy is a global property of execution, scheduling decisions are made locally and incrementally. 
Energy optimization poses different challenges compared to performance optimization in task-based systems.
While energy is an execution-wide, cumulative metric that depends on the collective effect of all scheduling decisions, task scheduling decisions are made online and sequentially as tasks become ready.
As a result, locally optimal choices, such as executing a task at the frequency that minimizes its standalone energy consumption, may increase global energy usage by delaying dependent tasks, prolonging idle periods on other GPUs, and ultimately extending the overall makespan, thereby offsetting per-task energy savings.
% These delays can significantly amplify system idle energy and offset per-task energy savings. 
For example, our characterization shows that an NVIDIA L4 GPU consumes approximately 27\,W when idle and waiting for tasks, accounting for nearly 38\% of its peak power (72\,W). % 
Moreover, task scheduling decisions are irreversible once dispatched, allowing early suboptimal choices to propagate through the dependency graph and constrain future decisions. 
Together, these factors make energy-aware scheduling in multi-GPU task graphs substantially more challenging.
% than performance-centric scheduling.

% While these multi-GPU architectures deliver exceptional computational throughput, achieving energy-efficient execution requires carefully orchestrating three interdependent factors: task placement across devices, Dynamic Voltage and Frequency Scaling (DVFS) configuration, and inter-GPU data movement.
% However, existing runtime systems lack integrated mechanisms to jointly optimize these factors, leaving significant energy savings unexploited or requiring substantial manual tuning by domain experts.

%%%%% DVFS is a powerful tool, determining the optimal one is not easy and it is not free of cost. %%%%%
% Dynamic Voltage and Frequency Scaling (DVFS) 
DVFS offers a compelling mechanism to improve energy efficiency by trading performance slack for reduced power consumption.
Modern GPUs expose fine-grained DVFS controls, enabling adaptive optimization based on workload heterogeneity.
% The potential of DVFS stems from workload heterogeneity: 
Different kernels exhibit distinct utilization patterns across GPU functional units,
% (compute cores, memory subsystem, interconnects)
making the optimal frequency highly kernel-dependent~\cite{GPUPowerModel2018, GPUDVFSSurvey2017, GPUFreqSelection2022}.
Prior work has demonstrated energy savings through DVFS-based scheduling, but primarily targets single-GPU execution~\cite{FineDVFS/ASPLOS2023, DNNGPUDVFS/ICPP2023, SYnergySC2023, MultiObjGPUKernSched2024, HuangGPUBalance2020, song2014energy, kernelmultitaskingRTSS2021, LUCPUGPUEnergy2023}.

In addition, frequency reconfiguration incurs non-trivial overhead, ranging from hundreds of microseconds to milliseconds depending on the transition magnitude and direction~\cite{GPUFreqSwitchLatency2025}.
This overhead makes fine-grained per-kernel tuning counterproductive for short-running tasks, while coarse-grained application-level policies often overlook substantial optimization opportunities by treating heterogeneous kernels uniformly.
To amortize this overhead, some prior work 
% avoids dynamic DVFS control altogether and instead 
adopts static or phase-based frequency selection strategies~\cite{PhaseBasedFS2025, PhaseDVFSModel/ICDCS2015}, which limits the ability to adapt to variable task granularity and dynamic runtime conditions.
A robust solution must make DVFS decisions online, determining when and how to adjust frequencies based on task granularity, computational characteristics, reconfiguration costs, and evolving system state within a unified optimization framework.

% Second, most prior work either ignores DVFS transition costs entirely or assumes simplified models (constant or linear overhead), while recent empirical studies reveal that GPU frequency transitions exhibit complex, asymmetric, and architecture-dependent behavior.

%%%%% Multi-GPU complexity %%%%%
The challenge intensifies in multi-GPU environments, where scheduling decisions span three interdependent dimensions.
First, \emph{frequency selection}: each GPU may operate at a different frequency, with optimal settings varying based on the task's computational characteristics and the current device state.
Second, \emph{task placement}: tasks must be mapped to GPUs considering not only device availability but also data locality and device-specific performance characteristics.
Third, \emph{cost awareness}: both inter-GPU data transfers and DVFS transitions impose latency and energy costs that can dominate savings for fine-grained operations.
These dimensions are inherently interdependent: the optimal frequency for a task depends on where it executes, placement must account for both DVFS reconfiguration and data movement costs, and data movement costs vary with both placement and device frequency.
This interdependence makes isolated optimization of any single dimension suboptimal, necessitating a holistic scheduling approach. 
However, prior work addresses these challenges largely in isolation. Approaches that focus on task placement target throughput improvement or load balancing across devices without integrating DVFS~\cite{MustardICS2015, CASEThroughputScheduling_2022, MultiGPUTaskPlace2024, MIGTaskPlace2025, TaskToSM/ICS2015}, while DVFS-focused techniques are commonly designed for single-device execution and do not account for the joint impact of task placement, inter-GPU data movement, and frequency transition overheads in multi-GPU systems.

To address the multi-dimensional challenges of energy-efficient task-graph execution on multi-GPU systems, we propose \textbf{DEFT} (\textbf{D}ynamic \textbf{E}nergy-aware \textbf{F}re\-quen\-cy -- \textbf{T}ask scheduling), a scheduler that jointly determines task-to-device placement and per-GPU DVFS configuration to achieve user-specified energy-performance trade-offs.
Users specify optimization objectives through a generalized energy-delay product metric ${ED^\beta P}$, where the exponent $\beta$ controls the relative importance of execution time versus energy.
% , $E^\alpha D^\beta P$ (EDP variant), where $\alpha$ and $\beta$ parametrize the relative importance of energy ($E$) and execution time ($D$).
% For example, $E^1D^1P$ (standard EDP) balances energy and performance equally, while ED$^2$P prioritizes performance, and E$^2$DP emphasizes energy reduction.
% We implement a prototype of DEFT on top of CUDASTF~\cite{CUDASTF2024}, a task-based CUDA runtime, extending its scheduling capabilities to incorporate DVFS control alongside task-to-device assignment.

% A key concept is task slack, defined as the difference between the application’s critical path length and the longest path passing through the task, which quantifies how much execution delay can be tolerated without extending overall makespan. 
% For each candidate configuration, DEFT predicts the task execution time and computes the induced delay relative to execution at the highest frequency, using slack to determine configuration feasibility. 
% Scheduling decisions follow a hierarchical policy: when comparing configurations across different GPUs, DEFT prioritizes the earliest feasible start time to preserve parallelism and avoid resource contention; when multiple slack-feasible configurations target the same GPU, DEFT selects the one that minimizes the specified $E^{\alpha}D^{\beta}P$, allowing frequency reduction to exploit available slack without impacting global progress.
DEFT adopts a two-phase decision strategy that hierarchically structures task placement and DVFS selection.
% In the first phase, the scheduler selects the device that yields the earliest feasible start time for each ready task, preserving task parallelism and reducing system-wide idle energy.
The first phase selects the execution device that yields the earliest feasible start time for each ready task, based on device availability and data readiness, and independently of DVFS, since frequency does not affect start-time feasibility.
In the second phase, DEFT determines the DVFS configuration on the selected device to optimize the energy--performance objective while preserving global progress.
DEFT integrates: 
(i) \emph{slack awareness}, derived from critical-path analysis, to bound execution delay without extending the application makespan;
(ii) \emph{throughput awareness} to prevent excessive slowdowns under resource contention when the number of ready tasks exceeds the available GPUs;
and (iii) \emph{cost awareness}, which explicitly models inter-GPU data transfer and DVFS transition overheads to avoid counterproductive decisions when these costs outweigh potential savings.
This unified design enables DEFT to adapt its optimization strategy to both task granularity and dynamic system conditions.
For short-lived tasks where reconfiguration or data movement overheads would dominate potential benefits, the scheduler favors local execution at the current frequency.
For longer-running or slack-rich tasks, DEFT explores a broader configuration space, trading execution time for improved energy efficiency.

In summary, this work makes the following contributions:
\begin{itemize} [leftmargin=*,itemsep=0pt,topsep=0pt] %
    \item \textbf{Energy-Aware Multi-GPU Scheduling Framework:} 
    % We present an energy-aware scheduling framework for task-graph execution on multi-GPU systems. The framework enables users to specify optimization objectives via generalized EDP metrics ($E^\alpha D^\beta P$) and automatically orchestrates task placement and DVFS configuration to achieve these objectives. While our prototype implementation targets CUDASTF, the methodology generalizes to other task-based runtimes.
    We prese\-nt a scheduling framework for task-graph execution on multi-GPU systems that enables users to specify optimization objectives and automatically orchestrates task placement and DVFS configuration. 
    % While our prototype targets CUDASTF, the methodology generalizes to other task-based runtimes.
    We prototype DEFT on top of the task-based CUDA runtime CUDASTF~\cite{CUDASTF2024}; however, the methodology generalizes to other task-based runtimes.
    
    % \item \textbf{Unified Cost-Model-Driven Scheduling Algorithm:} We develop a hierarchical scheduling algorithm that prioritizes device selection to preserve parallelism, then optimizes DVFS configuration on the selected device. 
    %  that jointly optimizes task-to-device assignment and per-GPU DVFS configuration. 
    % Across devices, the algorithm prioritizes earliest feasible start time to preserve parallelism; within the same device, it exploits feasible slack to tune frequency to minimize energy-delay objectives.
    % The algorithm explicitly models task execution cost, inter-GPU data movement, and DVFS transition overheads when evaluating configurations.

    \item \textbf{Cost-Model–Driven Hierarchical Scheduling Algorithm:} We develop a two-phase scheduling algorithm that prioritizes device selection to preserve parallelism and then optimizes the DVFS configuration on the selected device. The algorithm is guided by a unified cost model that explicitly captures task execution cost, inter-GPU data movement, DVFS transition overheads, and slack/throughput constraints when evaluating configurations. To our knowledge, DEFT is the first runtime framework to provide integrated, cost-aware coordination between placement and DVFS at task granularity in multi-GPU task-based runtimes.

    % \item \textbf{Predictive Models:} 
    % To support runtime scheduling decisions, we develop three complementary models for predicting task execution time, power consumption, and DVFS transition cost across the device-frequency configuration space.
    % We develop three predictive models that enable rapid, accurate runtime estimation for task execution time, power consumption, and DVFS transition cost across the device-frequency configuration space.
    % We develop three complementary models that enable rapid, accurate prediction: (1) an analytical power model that predicts tasks' average GPU power consumption based on component-level utilization; (2) an XGBoost-based performance model that predicts task execution time using hardware utilization; and (3) an empirical DVFS transition cost model that captures the asymmetric, non-linear, architecture-dependent overhead through piecewise functions.

    \item \textbf{Predictive Models:} We develop three complementary predictive models to support efficient runtime scheduling decisions across the device--frequency configuration space.
    Specifically, we (i) introduce an XGBoost-based performance model that predicts task execution time using hardware utilization features; (ii) adopt and extend an analytical power model that estimates task-level GPU power consumption based on component-level hardware utilization; and (iii) propose an empirical DVFS transition cost model that captures asymmetric and non-linear reconfiguration overheads through piecewise functions.

    % \item \textbf{Evaluation:} 
    % Experiments on NVIDIA L4 and L40S multi-GPU nodes using four scientific computing applications show that DEFT reduces energy by 14.8\% and 4.8\% on average on L40S and L4, respectively (9.9\% and 3.7\% in EDP), compared to native CUDASTF scheduling, while maintaining performance within 1.5\% of the fastest baseline.

    \item \textbf{Evaluation:}
    % We evaluate DEFT on NVIDIA L4 and L40S multi-GPU nodes using four scientific computing applications under five optimization objectives defined by the generalized metric $\mathbf{\mathrm{ED^{\beta}P}}$ (Energy-only, $\mathbf{\mathrm{E^2DP}}$, EDP, $\mathbf{\mathrm{ED^2P}}$, and Time-only).
    % Across applications and objectives, DEFT reduces energy by 14.8\% and 4.8\% on average on L40S and L4, respectively (9.9\% and 3.7\% in EDP), compared to native CUDASTF scheduling, while maintaining performance within 1.5\% of the fastest baseline.
    We evaluate DEFT on NVIDIA L4 and L40S multi-GPU nodes under five optimization objectives (Energy-only, $\mathbf{\mathrm{E^2DP}}$, EDP, $\mathbf{\mathrm{ED^2P}}$, and Time-only).
    DEFT reduces energy by 14.8\% and 4.8\% on average on L40S and L4, respectively (9.9\% and 3.7\% in EDP), compared to native CUDASTF scheduling, while maintaining performance within 1.5\% of the fastest baseline.

\end{itemize} 

% The remainder of this paper is organized as follows.
% Section~\ref{sec:relatedwork} surveys related work on energy-aware scheduling and task-based runtimes.
% Section~\ref{sec:methodology} describes the scheduling algorithm and implementation.
% Section~\ref{sec:models} details our performance and power modeling methodology.
% Section~\ref{sec:setup} outlines the experimental setup.
% Section~\ref{sec:evaluation} presents our experimental evaluation.
% Section~\ref{sec:conclusion} concludes the work. 

% DEFT employs a two-phase decision strategy: first selecting the device enabling the earliest feasible start time to preserve parallelism and reduce idle energy, then optimizing DVFS on the selected device to minimize the target objective.
% Central to DEFT is the concept of \emph{task slack}—derived from critical-path analysis—which quantifies how much execution delay a task can tolerate without extending the makespan.
% DEFT exploits slack to reduce frequency on non-critical tasks while remaining \emph{throughput-aware} (limiting execution-time inflation when ready tasks exceed available GPUs) and \emph{overhead-aware} (explicitly modeling data movement and DVFS transition costs).
% This enables adaptive behavior: short-lived tasks favor local execution at current frequency, while longer or non-critical tasks explore broader configuration spaces.
% To support real-time decisions, DEFT integrates predictive models for performance, power, and DVFS transition costs across the device-frequency spectrum.

\section{Background} \label{sec:background}

\begin{figure*}[!t]
    \centering
    % \vspace{-0.1cm}
    \includegraphics[width=\textwidth]{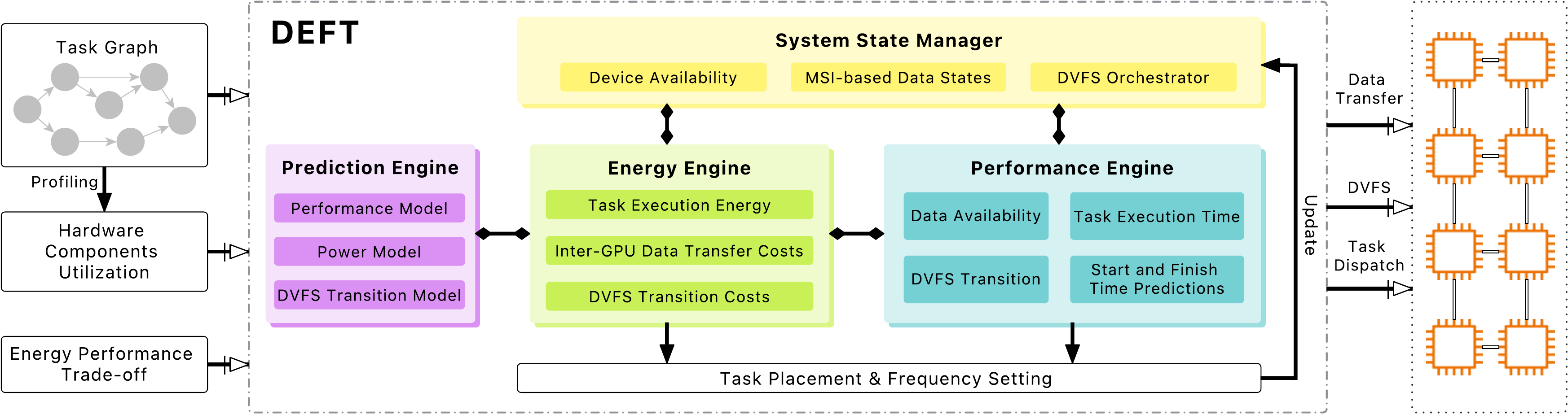}
    \vspace{-0.5cm}
    \caption{Overview of the proposed energy-aware scheduling framework.}
    % \vspace{-0.1cm}
    \label{fig:overview}
\end{figure*}

% This section introduces task-graph execution, scheduling terminology, and the CUDASTF runtime, which together form the foundation for the proposed scheduling framework.

\subsection{Task Graph}
Task-based parallel programming models provide a natural abstraction for expressing and optimizing dynamic parallelism in heterogeneous systems~\cite{openmp50-api, StarPU, cilk5, tbb_scheduler, Ompss2017}.
These models decompose applications into units of work (tasks) connected by data dependencies, forming a directed acyclic graph (DAG) that the runtime schedules automatically.
The runtime leverages schedulers to make resource allocation and hardware configuration decisions based on task characteristics, device capabilities, and optimization objectives.

In a DAG, the critical path is the longest dependency chain in the graph, which defines a lower bound on the application makespan, since any delay along this path directly extends total execution time. 
Tasks that do not lie on the critical path may tolerate limited execution delays without affecting the overall completion time. 
This tolerance is referred to as \emph{task slack}. 
Task slack can be derived through static analysis of the DAG by examining the longest paths from the entry to a task and from the task to an exit node under baseline execution assumptions. 
Conceptually, slack represents the degree of scheduling flexibility available for a task. 
This notion is especially important in runtime scheduling, as it enables the scheduler to distinguish between time-critical tasks that must be executed promptly and slack-rich tasks that can be delayed or slowed down to optimize secondary objectives, such as reducing energy consumption or avoiding resource contention.
% for energy-aware scheduling, as it allows the runtime to safely apply techniques such as DVFS to non-critical tasks trading execution speed for reduced energy consumption, while preserving progress along the critical path.

% This work leverages task-based programming models, which abstract applications into DAGs of data-dependent tasks and rely on the runtime system for automated scheduling and management.
% These models are essential as they allow developers to exploit increasingly complex machine organizations, such as heterogeneous hardware environments, by managing asynchronous operations that are difficult to coordinate manually.

\subsection{CUDASTF}
CUDASTF~\cite{CUDASTF2024} extends this task-based paradigm for multi-GPU CUDA applications and implements the Sequential Task Flow programming model, allowing applications to be expressed as a sequence of tasks that are automatically transformed into a DAG based on data dependencies inferred from task declarations.
% This approach is a form of dataflow that simplifies client source code by deriving dependencies rather than requiring them to be explicitly specified. 
The runtime automatically manages critical low-level operations, including data transfers, synchronization, and scheduling across multiple GPUs. 
A core abstraction is logical data,  which represents data as an abstract entity without associating it explicitly with specific storage locations; instead, it tracks multiple coherent replicas (data instances) across distinct physical memories. 
We leverage CUDASTF's task-graph abstraction and flexible design to transparently integrate energy-aware scheduling into task-based runtimes without application code modification.
CUDASTF supports both stream-based and CUDA Graph–based execution backends. 
% In this work, we build DEFT on top of the stream backend, which is better aligned with the dynamic, online scheduling and per-task DVFS decisions required by our approach.
We build DEFT on the stream backend, which enables the eager task execution and runtime flexibility necessary for per-task DVFS decisions and dynamic device assignment.
% . The stream backend's eager task execution model is essential for DEFT's dynamic scheduling, as it enables per-task DVFS configuration and runtime placement decisions based on current system state. In contrast,
% The graph backend's deferred execution and static parameter capture would prevent the runtime adaptability required for our approach.
% CUDASTF is particularly well-suited for our approach, because its automatic dependency inference eliminates the need for manual data movement specification and explicit graph declaration, reducing programmer burden, and its pluggable scheduler interface facilitates integration of custom optimization policies.
% (2) it can generate the dot file, enabling the task slack analysis (one time analysis) and visualization of task graphs;
% (3) its interception of kernel launches enables transparent profiling and scheduling; and (4) 

\section{Methodology} \label{sec:methodology}

% TODO: SAVE SPACE CHANGE - an overview of the proposed framwork
In this section, we first provide an overview of the proposed 
% energy-aware scheduling 
framework 
% for multi-GPU systems 
(\S~\ref{sec:method_overview}).
We then detail the scheduling algorithm and cost model (\S~\ref{sec:scheduling_algo}), followed by a discussion of the MSI-based coherence protocol for tracking data residency across GPUs (\S~\ref{sec:msi}).
% An illustrative scheduling example is provided next (Section~\ref{sec:example}).
Finally, we describe the profiling methodology for collecting per-task hardware utilization metrics (\S~\ref{sec:profiling}), which serve as inputs to the performance and power models (\S~\ref{sec:models}).

\subsection{Overview} \label{sec:method_overview}
% TODO: the first sentence can be skipped
% We present DEFT, an energy-aware scheduling framework that jointly optimizes task-to-device assignment and per-GPU DVFS configuration for task-graph execution on multi-GPU systems.
DEFT is guided by a unified cost model that accounts for task performance and power, inter-GPU data transfer, DVFS transition overheads, task slack, and throughput constraints.
% , enabling task-granularity–aware coordination of placement and frequency decisions.
Figure~\ref{fig:overview} illustrates the architectural overview of the DEFT framework.
It operates on three inputs: (1) an application expressed as a DAG with explicit data dependencies; 
(2) per-task hardware utilization metrics collected via lightweight one-time profiling (detailed in \S~\ref{sec:profiling}); (3) a user-specified optimization objective of the form $ED^\beta P$.
% , where $\alpha$ and $\beta$ parameterize the relative importance of energy reduction versus performance.
Our framework consists of four tightly coupled components that collaboratively make scheduling decisions.
Throughout this section, we use the following notation: $t_i$ denotes task $i$ in the task graph, $d$ denotes a GPU device index, and $f = (f_{\text{gpu}}, f_{\text{mem}})$ denotes a DVFS configuration comprising GPU core and memory frequencies.
The components are: % TODO: can be skipped
\begin{itemize} [leftmargin=*,itemsep=0pt,topsep=0pt]
    \item \textbf{Prediction Engine}: Comprises three models (detailed in \S~\ref{sec:models}) that predict task execution time $T_{i,d,f}$, average power consumption $P_{i,d,f}$, and DVFS reconfiguration latency, respectively.
    The performance and power models leverage hardware utilization metrics to extrapolate $T_{i,d,f}$ and $P_{i,d,f}$ across all device-frequency pairs $(d, f)$ without exhaustive runtime profiling.
    \item \textbf{Energy Engine}: Computes the total energy cost $E_{i,d,f}$ of scheduling task $t_i$ on device $d$ at frequency $f$, decomposed into three components: (i) task execution energy $E_{\text{exec}} = T_{i,d,f} \times P_{i,d,f}$, (ii) inter-GPU data transfer energy required to satisfy input dependencies, and (iii) DVFS reconfiguration energy incurred when transitioning device $d$ from its current frequency to $f$.
    \item \textbf{Performance Engine}: Computes task earliest start time $T_{\text{start}}$ and expected finish time $T_{\text{finish}}$ on each device by analyzing device and data availability, accounting for inter-GPU transfer latency, DVFS transition overhead, and task execution time.
    %  and (iii) updates the critical path length $D_{\text{critical}}$, which represents the projected makespan of the task graph after scheduling $t_i$. The critical path directly informs the delay component of the $E^\alpha D^\beta P$ objective.
    \item \textbf{System State Manager}: Maintains runtime state comprising (i) per-device availability times, (ii) MSI-based coherence metadata tracking data residency and validity across devices (described in \S~\ref{sec:msi}), and (iii) a DVFS orchestrator that checks for pending frequency change requests and idle GPU conditions. When a GPU remains idle beyond a predefined temporal threshold, the orchestrator proactively transitions it to a low-power state.
    % Overall, this manager enables context-aware scheduling decisions that adapt to evolving system conditions and task graph structure.
\end{itemize}

% For each ready task $t_i$ (i.e., all dependencies satisfied), the scheduler evaluates all feasible $(d, f)$ configurations by computing the projected cost according to Section~\ref{sec:scheduling_algo} and selects the configuration minimizing this objective.
Upon selecting configuration $(d^*, f^*)$, the System State Manager atomically updates: (i) device $d^*$'s workload and availability time, (ii) MSI coherence states for task $t_i$'s input and output data, and (iii) device $d^*$'s scheduled frequency to $f^*$.
The runtime subsequently initiates inter-GPU data transfers for missing dependencies, issues DVFS reconfiguration commands to set frequency $f^*$, and dispatches the task to device $d^*$ for execution.
% The detailed scheduling algorithm of how to leverage these components to achieve desired objectives is presented in Section~\ref{sec:scheduling_algo}.

\subsection{Scheduling Algorithm} \label{sec:scheduling_algo}
The scheduler processes tasks in topological order, ensuring all dependencies are satisfied before scheduling any dependent task.
For each ready task $t_i$, rather than exhaustively enumerating all $(d, f) \in \mathcal{D} \times \mathcal{F}$ pairs (where $\mathcal{D}$ is the set of available GPUs and $\mathcal{F}$ is the discrete DVFS configuration space), DEFT enables a hierarchical decision structure in which device selection is performed first based on task earliest start time, while DVFS configurations are evaluated conditionally on the selected device using the full cost model.
% we employ a hierarchical scheduling strategy that first selects the target device based on task earliest start time, and subsequently determines the operating frequency under a unified cost model.

% \textbf{Key insights.}
\subsubsection{Design Rationale} \label{sec:design_rationale} 
Prioritizing device selection based on earliest start time serves three critical objectives: (i) minimizing task completion time to unblock dependent tasks as early as possible, thereby preserving task-level parallelism in the DAG; (ii) reducing system-wide idle energy by avoiding unnecessary delays that force other devices to wait idly for data dependencies; and (iii) maintaining temporal locality to reduce inter-GPU communication overhead.
Crucially, the earliest time at which task $t_i$ can begin execution on device $d$ depends only on device availability, data availability, and the completion of its parent tasks—factors that are independent of DVFS configuration—enabling the two-phase decomposition. 
% This enables a two-phase decomposition: first select the device that minimizes $T_{\text{earliest}}^d$ (preserving parallelism and minimizing idle periods), then optimize the DVFS configuration for energy efficiency on that device (leveraging available slack without compromising the selected start time).

Scheduling decisions that achieve the lowest objective locally by aggressively reducing frequency may delay dependent tasks, prolong overall execution, and increase idle periods on other GPUs.
To bound this effect, we leverage task slack, 
% which quantifies the maximum execution delay a task can tolerate without extending the application makespan.
defined as the difference between the application’s critical path length and the longest path passing through the task, which quantifies how much execution delay can be tolerated without extending overall makespan. 
Task slack is precomputed via static analysis of the task graph using predicted execution times at maximum frequency.
When multiple tasks are ready, we prioritize tasks with lower slack to avoid extending the critical path and increasing overall makespan. 

However, dependency-aware slack alone is insufficient under resource contention, when task-level parallelism exceeds the number of available GPUs.
% However, dependency-aware slack alone is insufficient in resource constrained regimes, where task-level parallelism exceeds the number of available GPUs and resource contention becomes the dominant performance and energy efficiency limiter.
% While task slack bounds execution delay with respect to dependency constraints, it does not capture performance degradation caused by resource contention when task-level parallelism exceeds available GPUs. 
In such throughput-bound scenarios, aggressively slowing down slack-rich tasks risks delaying critical tasks, reducing overall system throughput, and prolonging the makespan despite preserving the critical path. 
We therefore monitor runtime contention by tracking the instantaneous number of ready tasks relative to the number of available GPUs, and activate a throughput-aware refinement when contention arises.
This refinement caps per-task execution-time inflation, preventing slack-rich tasks from monopolizing shared GPU resources and bounding slowdown to preserve overall throughput.

\subsubsection{Two-Phase Algorithmic Design} \label{sec:two_phase} 
\mbox{}\\
\textbf{Phase 1: Device Selection.}
For each device $d \in \mathcal{D}$, we first compute when task $t_i$'s input data will be available on device $d$:
\begin{equation} \label{eq:data_available}
T_{\text{data}}^d = \max_{p \in \text{parents}(t_i)} \left(T_{\text{finish}}^{p} + \frac{|{\text{data}}_p|}{BW}\right)
\end{equation}
where $T_{\text{finish}}^p$ is the scheduled finish time of parent task $p$, $|{\text{data}}_p|$ is the size of data produced by $p$ that $t_i$ requires, and $BW$ is the peer-to-peer bandwidth.
Inter-GPU transfers are handled by dedicated copy engines (e.g., NVLink or PCIe DMA) whose throughput is independent of GPU DVFS states and operates at fixed interconnect bandwidth.
If data resides on device $d$, the transfer time is zero. 
% (MSI state is Modified or Shared, as described in Section~\ref{sec:coherence})

The frequency-independent earliest start time for task $t_i$ on device $d$ is then:
\begin{equation} \label{eq:earliest_start}
T_{\text{earliest}}^d = \max\left(T_{\text{available}}^d, T_{\text{data}}^d\right)
\end{equation}
where $T_{\text{available}}^d$ is the time when device $d$ completes its currently scheduled workload.
This represents the earliest moment at which task $t_i$ can begin execution on device $d$, determined by the later of device availability or data readiness.

We also compute 
% auxiliary information for each device: the baseline execution time (at maximum frequency) $T_{i,d}^{\text{baseline}}$, and
the transfer energy $E_{\text{transfer}}^d$ in this phase.
When data arrives before the GPU becomes available ($T_{\text{data}}^d \leq T_{\text{available}}^d$), the GPU is busy and incurs no additional energy cost during the transfer (computation-transfer overlap).
When data arrives after the GPU is ready ($T_{\text{data}}^d > T_{\text{available}}^d$), the GPU idles during the transfer gap, consuming idle power:
\begin{equation}
E_{\text{transfer}}^d = \begin{cases}
0 & \text{if } T_{\text{data}}^d \leq T_{\text{available}}^d \\
(T_{\text{data}}^d - T_{\text{available}}^d) \times P_{\text{idle}}(f_{\text{curr}}^d) & \text{otherwise}
\end{cases}
\end{equation}
where $P_{\text{idle}}(f_{\text{curr}}^d)$ is the idle power of device $d$ at its current frequency configuration.

The device with the minimum $T_{\text{earliest}}^d$ is selected:
\begin{equation}
d^* = \argmin_{d \in \mathcal{D}} T_{\text{earliest}}^d
\end{equation}
% TODO: I think here can be skipped 
This selection prioritizes temporal availability, ensuring tasks are scheduled on the device that can start execution earliest, thereby minimizing data dependency stalls and critical path delays.

\textbf{Phase 2: Frequency Optimization.}
Having selected device $d^*$, we now evaluate all frequency configurations $f \in \mathcal{F}$ to find the best DVFS state.
For each candidate frequency $f = (f_{\text{gpu}}, f_{\text{mem}})$, we perform the following analysis:

\subsubsection*{\textnormal{Step \circlednum{1}: DVFS Transition Cost}}
% \noindent Step \circlednum{1}: DVFS Transition Cost.
The scheduler computes the cost of transitioning device $d^*$ from its current frequency configuration $f_{\text{curr}}$ to the candidate configuration $f$.
The transition latency $T_{\text{trans}}$ is predicted using the empirical DVFS transition model described in \S~\ref{sec:dvfs_model}.
During the transition, the device operates in an idle state, consuming:
\begin{equation}
E_{\text{trans}} = T_{\text{trans}} \times (P_{\text{idle}}(f_{\text{curr}}) + P_{\text{idle}}(f)) / 2
\end{equation}
where we approximate the average power during the transition as the mean of the idle power at the initial and target frequencies.

\subsubsection*{\textnormal{Step \circlednum{2}: Task Timing}}
% \noindent Step \circlednum{2}: Task Timing.
We assume that task $t_i$ cannot begin execution until both the DVFS transition completes and data is available.
The actual start time is:
\begin{equation} \label{eq:actual_start}
T_{\text{start}} = \max(T_{\text{available}}^{d^*} + T_{\text{trans}}, T_{\text{data}}^{d^*})
\end{equation}
Note that DVFS reconfiguration and data transfers occur in parallel on independent hardware (power management unit and copy engines, respectively); the start time is determined by whichever completes last.
The projected finish time is $T_{\text{finish}} = T_{\text{start}} + T_{i,d^*,f}$,
% \begin{equation}
% T_{\text{finish}} = T_{\text{start}} + T_{i,d^*,f}
% \end{equation}
where $T_{i,d^*,f}$ is the predicted execution time at frequency $f$ (from the Performance Model in \S~\ref{sec:performance_model}).

% For a task $t_i$, slack is defined based on static critical-path analysis of the task graph:
% \begin{equation}
% \text{Slack}(t_i) = D_{\text{CP}} - \left( R_{\text{up}}(t_i) + R_{\text{down}}(t_i) \right),
% \end{equation}
% where $D_{\text{CP}}$ denotes the length of the critical path of the DAG, $R_{\text{up}}(t_i)$ is the longest path from the entry node to $t_i$, and $R_{\text{down}}(t_i)$ is the longest path from $t_i$ to an exit node, both computed assuming baseline execution at the maximum frequency.

% Given a candidate frequency configuration $f$ on a selected device $d$, we compute the additional execution delay relative to the baseline as:
% \begin{equation}
% \Delta T(t_i, d, f) = T_{i,d,f} - T_{i,d}^{\text{baseline}}
% \end{equation}
% where $T_{i,d}^{\text{baseline}}$ is the predicted execution time at maximum frequency.
% A configuration is considered \emph{slack-feasible} if the induced delay can be absorbed by the task's available slack: $\Delta T(t_i, d, f) \leq \text{Slack}(t_i)$. 
% Configurations that violate this constraint are penalized in the cost function, discouraging frequency choices that would extend the global critical path.

\subsubsection*{\textnormal{Step \circlednum{3}: Slack Feasibility Analysis}}
% \noindent Step \circlednum{3}: Slack Feasibility Analysis.
Given a candidate frequency configuration $f$ on the selected device $d^*$, we first compute the additional execution delay relative to the baseline execution at maximum frequency:
\begin{equation}
\Delta T(t_i, d^*, f) = T_{i,d^*,f} - T_{i,d^*}^{\text{baseline}}
\end{equation}
% where $T_{i,d^*}^{\text{baseline}}$ denotes the predicted execution time at maximum frequency.
We define a configuration as \emph{slack-feasible} if the delay can be absorbed without extending the application makespan: $\Delta T(t_i, d^*, f) \leq \text{Slack}(t_i),$
% \begin{equation}
% \Delta T(t_i, d^*, f) \leq \text{Slack}(t_i),
% \end{equation}
where $\text{Slack}(t_i)$ represents the maximum tolerable delay of task $t_i$ derived from static critical-path analysis.

% While slack feasibility ensures critical-path preservation, it alone is insufficient when device-level parallelism is limited.
% When the number of concurrently ready tasks exceeds available GPUs, excessive frequency reduction on non-critical tasks may reduce system throughput, causing device queues to lengthen and ultimately increasing overall execution time and energy consumption.
% To prevent this, we incorporate a throughput-aware constraint that tightens slack budget used for frequency scaling under resource contention.
% % Specifically, when the runtime detects the number of ready tasks exceeds the number of available GPUs, we restrict the effective slack budget used for frequency scaling.
% We cap the DVFS-induced slowdown to a small fraction of the task’s baseline execution time, preventing excessive frequency reduction that would otherwise delay queued tasks and extend makespan.
% This mechanism discourages aggressive frequency reduction on non-critical tasks that contribute to sustaining device occupancy and parallel progress.
% Configurations that violate either slack or throughput feasibility are penalized in the cost function, ensuring that energy savings are pursued only when they do not compromise global execution progress.

When the number of ready tasks exceeds the number of available GPUs, we tighten the effective slack budget under resource contention by restricting the allowable DVFS-induced slowdown to the minimum of the task slack and a small fraction of its baseline execution time:
% , preventing excessive frequency reduction that would delay queued tasks and degrade global progress. 
\begin{equation}
\Delta T(t_i, d^*, f) \leq \min{\{\text{Slack}(t_i), \kappa \times T_{i,d^*}^{\text{baseline}}\}}
\end{equation}
This constraint jointly enforces slack feasibility and throughput preservation under contention.
Configurations that violate either bound are considered infeasible and penalized in the cost function.
In our experiments we use a fixed, conservative cap of $\kappa=2\%$, selected via preliminary sensitivity testing to bound per-task slowdown under contention.
% Since $\kappa$ is only active under resource pressure, the qualitative conclusions are insensitive to fine-tuning within this small range.

\subsubsection*{\textnormal{Step \circlednum{4}: Energy Cost Evaluation}}
% \noindent Step \circlednum{4}: Energy Cost Evaluation.
The execution energy for task $t_i$ at configuration $f$ is: 
$E_{\text{exec}} = T_{i,d^*,f} \times P_{i,d^*,f}$.
% \begin{equation}
% E_{\text{exec}} = T_{i,d^*,f} \times P_{i,d^*,f}
% \end{equation}
% where $P_{i,d^*,f}$ is the average power predicted by the Power Model (Section~\ref{sec:power_model}).
The total incremental energy cost is
%  $E_{i,d^*,f} = E_{\text{exec}} + E_{\text{trans}} + E_{\text{transfer}}^{d^*}$.
\begin{equation}
E_{i,d^*,f} = E_{\text{exec}} + E_{\text{trans}} + E_{\text{transfer}}^{d^*}
\end{equation}

% For slack-feasible configurations, we use energy as the base cost. For slack-infeasible configurations, we add a large penalty proportional to the slack violation:
% \begin{equation}
% \text{BaseCost}(f) = \begin{cases}
% \Delta E_{i,d^*,f} & \text{if } \Delta T \leq \text{Slack}(t_i) \\
% \Delta E_{i,d^*,f} + \lambda \cdot (\Delta T - \text{Slack}(t_i)) & \text{otherwise}
% \end{cases}
% \end{equation}
% where $\lambda$ is a large penalty constant (e.g., $10^6$) that effectively discourages critical path extension.

\subsubsection*{\textnormal{Step \circlednum{5}: Configuration Selection}}
% \noindent Step \circlednum{5}: Configuration Selection.
The scheduler selects the frequency that minimizes a generalized cost function:
\begin{equation}
\label{eq:cost_function}
f^* = \argmin_{f \in \mathcal{F}} \left[E_{i,d^*,f} \times (T_{i,d^*,f})^{\beta}\right]
\end{equation}
% where $\alpha, \beta \geq 0$ are user-specified exponents controlling the energy-performance trade-off.
% \textbf{Hierarchical Comparison Logic.}
% When comparing frequency candidates, the scheduler employs a hierarchical decision rule:
% \begin{enumerate}
%     \item \textbf{Slack feasibility dominates:} Slack-feasible configurations always outrank slack-infeasible ones.
%     \item \textbf{Among slack-feasible configurations:}
%     \begin{itemize}
%         \item If the delay from $T_{\text{earliest}}^{d^*}$ to $T_{\text{start}}$ is solely due to DVFS transition (i.e., $T_{\text{start}} - T_{\text{earliest}}^{d^*} \approx T_{\text{trans}}$), prioritize \emph{energy} (lower BaseCost).
%         \item Otherwise (data transfer delay dominates), prioritize \emph{start time}, using energy as a tie-breaker.
%     \end{itemize}
%     \item \textbf{Among slack-infeasible configurations:} Select by lowest cost (Equation~\ref{eq:cost_function}).
% \end{enumerate}
% This logic ensures that when the device must wait for data, we prefer configurations that start execution sooner (reducing idle time and energy cost), while when the device is already ready, we prefer energy-efficient frequencies that leverage available slack.

\subsubsection*{\textnormal{Step \circlednum{6}: State Update}}
% \noindent Step \circlednum{6}: State Update.
Upon selecting $(d^*, f^*)$, the System State Manager performs the following updates: (i) $T_{\text{available}}^{d^*} \leftarrow T_{\text{finish}}$; (ii) MSI coherence states are updated to reflect that $t_i$'s output data will be in Modified state on device $d^*$ upon completion; (iii) device $d^*$'s scheduled frequency set to $f^*$.
% \begin{enumerate}
%     \item \textbf{Device availability:} $T_{\text{available}}^{d^*} \leftarrow T_{\text{finish}}$
%     % \item \textbf{Critical path:} $D_{\text{critical}} \leftarrow \max(D_{\text{critical}}^{\text{current}}, T_{\text{finish}})$
%     \item \textbf{Data residency:} MSI coherence states updated to reflect that $t_i$'s output data will be in Modified state on device $d^*$ upon completion
%     \item \textbf{Frequency state:} Device $d^*$'s scheduled frequency set to $f^*$
% \end{enumerate}

The two-phase process repeats for each subsequent task in topological order until the entire task graph is scheduled.
We describe DEFT in the context of multi-GPU systems; however, the same scheduling procedure applies to single-GPU execution, where device selection is degenerate and only DVFS decisions remain.

\subsection{MSI-Based Data Residency Tracking} \label{sec:msi}

To accurately model data transfer and avoid unnecessary inter-GPU communication, we maintain per-data-instance coherence state using a protocol inspired by the Modified-Shared-Invalid (MSI) cache coherence abstraction~\cite{CUDASTF2024}.
Unlike hardware MSI implementations that operate at cache-line granularity with synchronous runtime invalidations, the protocol operates at task-output granularity and maintains logical coherence through scheduler-driven state updates at scheduling time.
Each logical data object (tasks' inputs and outputs) is associated with one of three per-device coherence states: (1) \textbf{Modified (M):} Device $d$ holds the exclusive, up-to-date copy; all other devices are Invalid. 
(2) \textbf{Shared (S):} Device $d$ holds a valid, read-only copy; other devices may also hold Shared copies.
(3) \textbf{Invalid (I):} Device $d$ does not have a valid copy; data must be transferred before use.
% \begin{itemize} [leftmargin=*,itemsep=0pt,topsep=0pt] 
%     \item \textbf{Modified (M):} Device $d$ holds the exclusive, up-to-date copy; all other devices are Invalid.
%     \item \textbf{Shared (S):} Device $d$ holds a valid, read-only copy; other devices may also hold Shared copies.
%     \item \textbf{Invalid (I):} Device $d$ does not have a valid copy; data must be transferred before use.
% \end{itemize}

\begin{figure}[!t]
    \centering
    \includegraphics[width=\columnwidth]{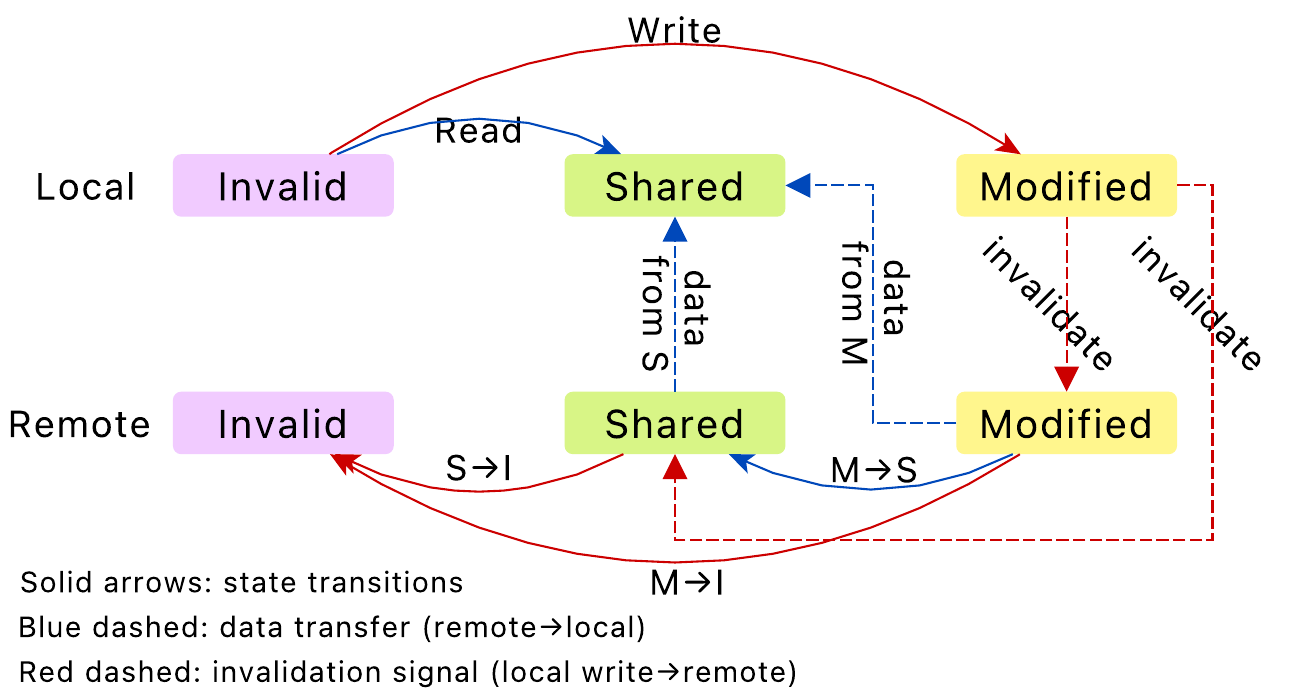}
    \vspace{-0.5cm}
    \caption{MSI coherence protocol for data instance state transitions in local and remote devices. 
    % Read operations transfer data from remote devices (blue dashed arrows) and downgrade remote Modified to Shared. Write operations transition local state to Modified and send invalidation signals (red dashed arrows) to remote devices.
    }
    \vspace{-0.4cm}
    \label{fig:msi_transitions}
\end{figure}

 % adhere to standard MSI semantics, as summarized in 
Figure~\ref{fig:msi_transitions} summarizes the state transitions and data flow.
When task $t_i$ executes on device $d$ and produces output $o$:
\begin{itemize} [leftmargin=*,itemsep=0pt,topsep=0pt] %
    \item Write (produce): Device $d$'s state for $o$ transitions to Modified; all other devices' states transition to Invalid.
    \item Read (consume): If device $d$ has state Invalid, it transitions to Shared and data is transferred from a device in state Modified or Shared; the source device’s state remains Shared or is downgraded from Modified to Shared.
\end{itemize}
% ORIGINAL: Remote device states are updated accordingly: a remote \emph{Modified} state is downgraded to \emph{Shared} upon a read on device~$d$ or \emph{Invalid} upon a remote write, while any \emph{Shared} state is invalidated when a write occurs elsewhere.
This protocol allows DEFT to maintain an up-to-date logical view of data residency and coherence across devices, enabling determination of data transfer requirements via simple state queries and accurate estimation of inter-GPU data transfer latency and energy.
% This protocol ensures that: (i) at most one device holds Modified state per data object (single-writer invariant), (ii) multiple devices may simultaneously hold Shared state (multi-reader), and (iii) the scheduler can determine data transfer requirements via simple state queries without runtime synchronization overhead.

\begin{table*}[t]
\centering
\vspace{-0.2cm}
\caption{Methodology for calculating hardware utilization ratios and required NCU metrics (L40S GPU example).}
\vspace{-0.3cm}
\label{tab:ncu_methodology}
\resizebox{\textwidth}{!}{%
\begin{tabular}{@{}lllc@{}}
\toprule
\textbf{Component} & \textbf{Utilization Formula / Method} & \textbf{NCU Metrics Required} & \textbf{\begin{tabular}[c]{@{}c@{}}Peak Rate Constant \end{tabular}} \\ \midrule
INT32 (ALU) &
  $\dfrac{\text{smsp\_\_thread\_inst\_executed\_pipe\_alu\_pred\_on.sum}}{\text{smsp\_\_cycles\_active.sum} \times \text{Peak\_Rate}_{\text{ALU}}}$ &
  \begin{tabular}[l]{@{}l@{}}\texttt{smsp\_\_thread\_inst\_executed\_pipe\_alu\_pred\_on.sum}\\ \texttt{smsp\_\_cycles\_active.sum}\end{tabular} &
  16 inst/cycle/SMSP \\ \addlinespace
FP32 (FMA) &
  $\dfrac{\text{smsp\_\_thread\_inst\_executed\_pipe\_fma\_pred\_on.sum}}{\text{smsp\_\_cycles\_active.sum} \times \text{Peak\_Rate}_{\text{FMA}}}$ &
  \begin{tabular}[l]{@{}l@{}}\texttt{smsp\_\_thread\_inst\_executed\_pipe\_fma\_pred\_on.sum}\\ \texttt{smsp\_\_cycles\_active.sum}\end{tabular} &
  32 inst/cycle/SMSP \\ \addlinespace
FP16 &
  $\dfrac{\text{smsp\_\_thread\_inst\_executed\_pipe\_fp16\_pred\_on.sum}}{\text{smsp\_\_cycles\_active.sum} \times \text{Peak\_Rate}_{\text{FP16}}}$ &
  \begin{tabular}[l]{@{}l@{}}\texttt{smsp\_\_thread\_inst\_executed\_pipe\_fp16\_pred\_on.sum}\\ \texttt{smsp\_\_cycles\_active.sum}\end{tabular} &
  32 inst/cycle/SMSP \\ \addlinespace
FP64 &
  $\dfrac{\text{smsp\_\_thread\_inst\_executed\_pipe\_fp64\_pred\_on.sum}}{\text{smsp\_\_cycles\_active.sum} \times \text{Peak\_Rate}_{\text{FP64}}}$ &
  \begin{tabular}[l]{@{}l@{}}\texttt{smsp\_\_thread\_inst\_executed\_pipe\_fp64\_pred\_on.sum}\\ \texttt{smsp\_\_cycles\_active.sum}\end{tabular} &
  0.5 inst/cycle/SMSP \\ \addlinespace
SFU (XU) &
  $\dfrac{\text{smsp\_\_thread\_inst\_executed\_pipe\_xu\_pred\_on.sum}}{\text{smsp\_\_cycles\_active.sum} \times \text{Peak\_Rate}_{\text{XU}}}$ &
  \begin{tabular}[l]{@{}l@{}}\texttt{smsp\_\_thread\_inst\_executed\_pipe\_xu\_pred\_on.sum}\\ \texttt{smsp\_\_cycles\_active.sum}\end{tabular} &
  4 inst/cycle/SMSP \\ \addlinespace
Tensor (Half) &
  $\dfrac{\text{smsp\_\_pipe\_tensor\_op\_hmma\_cycles\_active.sum}}{\text{smsp\_\_cycles\_active.sum} }$ &
  \begin{tabular}[l]{@{}l@{}}\texttt{smsp\_\_pipe\_tensor\_op\_hmma\_cycles\_active.sum}\\ \texttt{smsp\_\_cycles\_active.sum}\end{tabular} &
  N/A \\ \addlinespace
Tensor (Int) &
  $\dfrac{\text{smsp\_\_pipe\_tensor\_op\_imma\_cycles\_active.sum}}{\text{smsp\_\_cycles\_active.sum} }$ &
  \begin{tabular}[l]{@{}l@{}}\texttt{smsp\_\_pipe\_tensor\_op\_imma\_cycles\_active.sum}\\ \texttt{smsp\_\_cycles\_active.sum}\end{tabular} &
  N/A \\ \addlinespace \midrule
% DRAM &
%   \begin{tabular}[l]{@{}l@{}}$\dfrac{\text{Achieved Bandwidth}}{\text{Peak Bandwidth}}$, where \\ Achieved Bandwidth = $\dfrac{\text{dram\_\_bytes.sum}}{\text{gpu\_\_time\_duration.sum}}$\end{tabular} &
%   \begin{tabular}[l]{@{}l@{}}\texttt{dram\_\_bytes.sum}\\ \texttt{gpu\_\_time\_duration.sum}\end{tabular} &
%   864 GB/s \\ \addlinespace
DRAM &
  $\dfrac{\text{Achieved BW}}{\text{Peak BW}}$, where Achieved BW = $\dfrac{\text{dram\_\_bytes.sum}}{\text{gpu\_\_time\_duration.sum}}$ &
  \begin{tabular}[l]{@{}l@{}}\texttt{dram\_\_bytes.sum}\\ \texttt{gpu\_\_time\_duration.sum}\end{tabular} &
  864 GB/s \\ \addlinespace
L2 Cache &
  $\dfrac{\text{Achieved BW}}{\text{Peak BW}}$ where Achieved BW = $\dfrac{\texttt{lts\_\_t\_bytes.sum}}{\texttt{gpu\_\_time\_duration.sum}}$ &
  \begin{tabular}[l]{@{}l@{}}\texttt{lts\_\_t\_bytes.sum}\\ \texttt{gpu\_\_time\_duration.sum}\end{tabular} &
  45740 GB/s \\ \addlinespace
L1 Cache &
  $\dfrac{\text{l1tex\_\_t\_sector\_hit\_rate.pct}}{100}$ (Hit Rate) &
  \texttt{l1tex\_\_t\_sector\_hit\_rate.pct} &
  N/A \\ \addlinespace
Shared Memory &
  $\dfrac{\text{smsp\_\_sass\_l1tex\_pipe\_lsu\_wavefronts\_mem\_shared.sum}}{\text{l1tex\_\_data\_pipe\_lsu\_wavefronts.sum}}$ &
  \begin{tabular}[l]{@{}l@{}}\texttt{smsp\_\_sass\_l1tex\_pipe\_lsu\_wavefronts\_mem\_shared.sum}\\ \texttt{l1tex\_\_data\_pipe\_lsu\_wavefronts.sum}\end{tabular} &
  N/A \\ \bottomrule
\end{tabular}%
}
\end{table*}

% Why "lightweight profiling"? 
% - Profiling happens ONCE per unique kernel (not continuously)
% - Occurs at first encounter (on-demand, not offline)
% - Uses NCU to collect metrics 
% - Subsequent invocations use cached data (no re-profiling)

\subsection{Lightweight One-Time Profiling} \label{sec:profiling}
DEFT relies on accurate task performance and power predictions to guide scheduling decisions.
The Prediction Engine leverages models trained on hardware utilization metrics to estimate execution time and power consumption for each task across all $(d, f) \in \mathcal{D} \times \mathcal{F}$.
% To predict performance and power consumption across DVFS configurations, the scheduler leverages models that require hardware utilization characteristics for each task.
Rather than exhaustive offline profiling or continuous runtime monitoring, we employ a lightweight profiling approach based on a single Nsight Compute~\cite{NsightCompute} characterization run.
We execute the application once and collect a fixed set of hardware performance counters for all target kernels at the default frequency, using a bounded number of launches per kernel to limit profiling overhead. 
% This single profiling run provides representative utilization statistics for each kernel instance encountered in the application.

The profiling captures utilization rates across major GPU functional units, as described in \S~\ref{sec:training_data} and summarized in Table~\ref{tab:ncu_methodology}. 
% These rates include compute pipeline utilization (INT32, FP32, FP16, FP64, SFU, Tensor cores) and memory subsystem utilization (DRAM, L2, L1, and shared memory).
Each metric is normalized as a fraction of the hardware’s theoretical peak throughput.
The resulting utilization vectors serve as input to the Prediction Engine.

% The profiling captures detailed utilization statistics across all major GPU functional units, as described in Section~\ref{sec:training_data} and summarized in Table~\ref{tab:ncu_methodology}.
% These metrics include compute pipeline utilization (INT32, FP32, FP16, FP64, SFU, Tensor cores) and memory subsystem utilization (DRAM, L2 cache, L1 cache, shared memory).
% Each metric is normalized to represent a utilization fraction relative to the hardware's theoretical peak throughput.
% The collected utilization numbers become the input to the performance and power models in the Prediction Engine.
% The power model uses them to predict GPU power consumption $P_{i,d,f}$ at any DVFS configuration by decomposing power across functional units.
% The performance model uses them to predict speedup $S_{i,d,f}$ relative to default frequency execution.
% Together, these enable prediction of $(T_{i,d,f}, P_{i,d,f})$ across all $(d, f) \in \mathcal{D} \times \mathcal{F}$ from a single profiling run.
In most cases, a task corresponds to an invocation of a single kernel, 
and the collected kernel-level metrics are directly used to characterize the task. 
% and the collected kernel-level metrics directly characterize the task.
In cases where a task comprises multiple kernel launches, we aggregate the kernel-level metrics using an execution-time weighted average to derive a single utilization vector that represents the task’s overall hardware behavior.
% Because profiling is performed once for the entire application and only for a bounded number of kernel launches, the overhead is on the order of minutes and amortized across all subsequent executions.
Because profiling is performed once per application and on a bounded number of launches per kernel, the cost scales with the number of distinct task types and profiled launches rather than DAG size, is on the order of minutes for our benchmarks, and is amortized across subsequent executions.
% We leave the treatment of applications with highly diverse task types or dynamically generated task graphs to future work.
We leave applications with input-dependent task behavior or dynamically generated task graphs to future work.

\section{Models} \label{sec:models}
This section describes the three models comprising the Prediction Engine.
We first present the offline profiling methodology used to construct the training datasets (\S~\ref{sec:training_data}), then detail the power model and performance model that predict task average power consumption and execution time across DVFS configurations (\S~\ref{sec:power_model}--\ref{sec:performance_model}).
Finally, we describe the DVFS transition cost model that quantifies frequency reconfiguration overhead (\S~\ref{sec:dvfs_model}).

\subsection{Training Dataset Generation} \label{sec:training_data}
% ORIGINAL: To accurately model the task characteristics on underlying architectures, the proposed power and performance models rely on the profiled utilization ratio of the modeled GPU components.
To capture task behavior across DVFS configurations, our power and performance models require hardware utilization metrics that characterize workload demands on GPU functional units.
% ORIGINAL: The methodology for generating the training dataset follows the characterization approach introduced in prior work~\cite{GPUPowerModel2018, GPUPowerDVFS2019}, with modifications tailored to our scheduler and modern GPU architectures.
We adapt the microbenchmark-based characterization methodology from prior work~\cite{GPUPowerModel2018, GPUPowerDVFS2019}, extending it to support modern GPU architectures and the runtime prediction requirements of our scheduler.
% ORIGINAL: The training dataset is built by developing a suite of custom microbenchmarks, each carefully designed to stress a specific GPU hardware component to different extent.
The training dataset is constructed using a suite of synthetic microbenchmarks (244 in total), each designed to exercise specific GPU functional units, including INT32 (ALU), FP32 (FMA), FP16, FP64, SFU (special functions), shared memory, L1 cache, L2 cache, DRAM, and Tensor cores (FP16/FP8 and INT8/INT4), at varying intensities.
Specifically, we control the arithmetic intensity of each microbenchmark by changing the number of compute or memory operations performed within tight loops, allowing us to generate a wide spectrum of utilization values for each hardware component.

% \subsubsection{Profiling Methodology}
We begin by establishing a baseline hardware profile for each microbenchmark.
Each benchmark executes at the default frequency configuration ($f_{\text{gpu}}^{\text{default}}, f_{\text{mem}}^{\text{default}}$) while we collect detailed hardware performance counters via NVIDIA Nsight Compute (NCU).
% ORIGINAL: The collected metrics (summarized in Table~\ref{tab:ncu_methodology}) provide a detailed signature of the kernel's intrinsic resource demands.
These metrics (shown in Table~\ref{tab:ncu_methodology}) capture a kernel's resource consumption across the major GPU functional units, forming a hardware utilization signature independent of frequency scaling.
% ORIGINAL: From these raw values, we derive a normalized utilization vector that reflects the fraction of peak capacity consumed by each major hardware component.
From these raw counter values, we compute a normalized utilization vector $U = [U_j]$, where each component $U_j \in [0, 1]$ represents the fraction of peak theoretical throughput achieved by $j$.
% ORIGINAL: For example, the utilization of the integer (INT) compute pipeline is computed by dividing the total number of integer thread instructions executed by the theoretical maximum number of instructions that could have been executed in the measured active cycles.
For example, integer (INT) pipeline utilization is computed using the number of executed integer ALU instructions: 
\begin{equation}
U_{\text{INT}} = \frac{\text{ALU Instructions Executed}}{\text{Active Cycles} \times \text{Peak\_Rate}_{\text{ALU}}},
\end{equation}
where $\text{Peak\_Rate}_{\text{ALU}}$ is the architecture-specific throughput (e.g.,~16 instructions/cycle/SM partition for L40S, from Table~\ref{tab:ncu_methodology}~\cite{AdaWhitePaper}).
We compute memory subsystem utilization by normalizing measured throughput against the theoretical peak bandwidth. 
DRAM theoretical peak bandwidth $\text{BW}_{\text{DRAM}}$ is derived from the external memory interface specifications: 
$\text{BW}_{\text{DRAM}} = (\text{R}_{\text{transfer}} \times \text{W}_{\text{bus}})/8$, 
% \begin{equation} \label{eq:dram_bw}
% % \text{BW}_{\text{DRAM}} = \frac{\text{effective transfer rate (MT/s)} \times \text{Bus Width (bits)}}{8 \text{ (bits/byte)}}
% \text{BW}_{\text{DRAM}} = \frac{\text{R}_{\text{transfer}} \times \text{W}_{\text{bus}}}{8}
% \end{equation}
where $\text{R}_{\text{transfer}}$ is the effective transfer rate (MT/s) and $\text{W}_{\text{bus}}$ is the aggregate memory interface width (in bits).
L2 cache peak bandwidth $\text{BW}_{\text{L2}}$ represents the aggregate throughput of the on-chip interconnect. 
It is calculated as the product of the data path width per SM, the total number of SMs, and the GPU's maximum graphics clock:
% It is calculated by multiplying the bandwidth of a single Streaming Multiprocessor's (SM) connection to the L2 fabric by the total number of SMs on the device and the GPU's boost clock frequency:
\begin{equation} \label{eq:l2_bw}
\text{BW}_{\text{L2}} = \left( \frac{\text{Bytes}}{\text{Cycle} \times \text{SM}} \right) \times \text{N}_{\text{SMs}} \times f_{\text{gpu}}^{\text{max}}.
\end{equation}
% These architecturally-grounded theoretical maximums provide a consistent and reproducible baseline for normalizing the measured memory traffic into the final utilization ratios used in our model.
% where $N_{\text{SMs}}$ is the total number of SMs and $f_{\text{gpu}}^{\text{boost}}$ is the boost clock frequency. 
% These architectural constants provide device-specific normalization factors for memory utilization metrics.

% \subsubsection{Dataset Construction}
We execute each microbenchmark across the full spectrum of supported $(f_{\text{gpu}}, f_{\text{mem}})$, recording the execution time and average steady-state power $P$. 
These measurements are combined with the baseline utilization vector $U$ (profiled at the default frequency) to generate training tuples: $(U, f_{\text{gpu}}, f_{\text{mem}}, P)$ for the power model and $(U, f_{\text{gpu}}, f_{\text{mem}}, S)$ for the performance model, where $S$ is the speedup factor relative to the execution time at the default frequency.
% For each configuration, we: (i) set GPU frequencies via NVML~\cite{NVML2025}, (ii) execute the benchmark, (iii) sample instantaneous power readings via NVML throughout execution, and (iv) compute average steady-state power consumption $P$.
% For each configuration, we set the GPU frequencies, execute the benchmark, sample instantaneous power readings throughout execution, record the execution time, and compute average steady-state power consumption $P$.
% The power model training dataset comprises tuples $(U, f_{\text{gpu}}, f_{\text{mem}}, P)$,
% where $U$ is the utilization vector measured once at default frequency, and $P$ is the average power at $(f_{\text{gpu}}, f_{\text{mem}})$.
% Similarly, the performance model training dataset comprises tuples $(U, f_{\text{gpu}}, f_{\text{mem}}, S)$, where $S = T_{f_{\text{gpu}}, f_{\text{mem}}} / T_{\text{default}} $ is the speedup factor relative to execution time $T_{\text{default}}$ at the default frequency.

\begin{figure*}[!t]
    \centering
    \vspace{-0.3cm}
    \includegraphics[width=\textwidth]{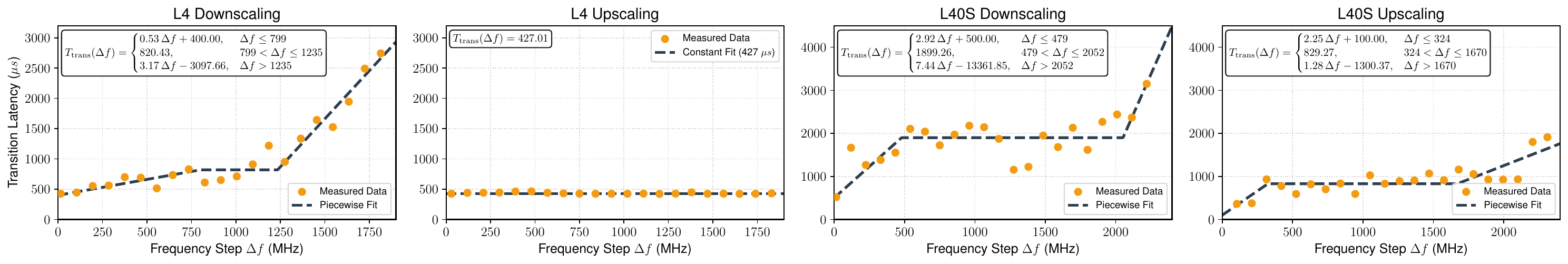}
    \vspace{-0.7cm}
    \caption{Measured DVFS transition latencies on NVIDIA L4 and L40S GPUs, showing asymmetric and non-monotonic behavior.}
    \vspace{-0.3cm}
    % modeled using multi-tier piecewise functions.}
    \label{fig:dvfs_latency}
\end{figure*}

\subsection{Power Model} \label{sec:power_model}
% ORIGINAL: To accurately predict GPU power consumption under different DVFS settings, we employ an analytical model based on the methodology proposed in the prior work~\cite{GPUPowerModel2018, GPUPowerDVFS2019} with modifications tailored to our scheduler.
% REVISED: More direct and specific
We employ an analytical power model adapted from prior work~\cite{GPUPowerModel2018, GPUPowerDVFS2019} to predict GPU power consumption $P(U, f_{\text{gpu}}, f_{\text{mem}})$ given workload utilization $U$ and DVFS configuration.
% ORIGINAL: The proposed power model is based on the utilization ratio of the modeled GPU components.
% DELETED: Redundant with introduction
% ORIGINAL: This approach models the total GPU power as the sum of static, constant dynamic (idle), and utilization-dependent dynamic power components.
% REVISED: More technical
The model decomposes total GPU power into three components: static leakage, frequency-dependent idle power, and utilization-dependent dynamic power.
% ORIGINAL: The model is defined by the following generalized equation:
% REVISED: More direct
Formally, for frequency domains $k \in \{\text{core}, \text{memory}\}$:
\vspace{-0.1em}
\begin{equation}
\begin{split}
P_{\text{total}} = \textstyle{\sum_{k \in \text{Domains}}} ( &\gamma_{\text{static},k} \cdot V_k + \gamma_{\text{idle},k} \cdot f_k \cdot V_k^2 \\
&+ \textstyle{\sum_{j \in C_{k}}} \omega_{j,k} \cdot U_j \cdot f_k \cdot V_k^2 )
\end{split} 
\end{equation}
% \begin{equation}
% P_{\text{total}} =
% \sum\limits_{k\in\text{Domains}}
% \!\left(
% \gamma_{\text{static},k}V_k
% +\gamma_{\text{idle},k}f_kV_k^2
% +\sum\limits_{j\in C_k}\!\omega_{j,k}U_j f_k V_k^2
% \right)
% \end{equation}
where $f_k$ and $V_k$ are the clock frequency and normalized supply voltage for domain $k$, and $C_k$ denotes the set of functional units in domain $k$.
% , and $U_j \in [0, 1]$ is the normalized utilization of unit $j$.
The model parameters are: $\gamma_{\text{static},k}$ and $\gamma_{\text{idle},k}$ (static and idle power coefficients), and $\omega_{j,k}$ (dynamic power coefficient per unit utilization for component $j$).

% \textbf{Training Procedure.}
In the proposed power model, the voltage–frequency relationships and power-model coefficients are jointly estimated through an iterative optimization procedure that alternates between two steps. 
% \textbf{(i) Voltage Estimation:}
Given current coefficient estimates $\{\gamma, \omega\}$, the supply voltage $V_k$ at each frequency $f_k$ is obtained by treating the power equation as a nonlinear system in $V_k$.
We apply constrained nonlinear least-squares regression, enforcing monotonic dependence of $V_k$ on $f_k$, to fit voltage values that minimize prediction error over the training set.
This yields a normalized voltage-frequency (V-F) curve satisfying $V_k(f_{\text{ref}}) = 1.0$ for a reference frequency $f_{\text{ref}}$.
% \textbf{(ii) Coefficient Estimation:}
With $V_k(f_k)$ fixed, the power equation becomes linear in coefficients $\{\gamma, \omega\}$.
Non-negative least-squares regression is then used to estimate the coefficients that minimize prediction error.
These two steps are repeated iteratively until convergence (i.e.,~coefficient updates fall below a predefined threshold).
%  (i.e., $\|\Delta\gamma\| < \epsilon$, $\|\Delta\omega\| < \epsilon$).
The converged parameters define the final power model $P(U, f_{\text{gpu}}, f_{\text{mem}})$ used for runtime prediction.

\subsection{Performance Model} \label{sec:performance_model}
Unlike power consumption, task execution time exhibits complex, non-linear dependencies on multiple factors, such as task characteristics, instruction throughput, memory bandwidth, latency hiding efficiency, and inter-component contention.
These non-linear interactions preclude simple analytical formulations, motivating a machine-learning-based approach.
An XGBoost gradient-boosted decision tree regressor is employed to learn the mapping $(U, f_{\text{gpu}}, f_{\text{mem}}) \rightarrow S$, where $S$ is the speedup factor relative to execution at the default frequency.
XGBoost is well suited to this task due to its ability to capture non-linear feature interactions through tree ensembles, handle high-dimensional utilization features efficiently, and provide robust predictions with limited training data~\cite{XGBoost2016}.

The model takes as input the normalized utilization metrics from Table~\ref{tab:ncu_methodology} along with the normalized GPU and memory frequency ratios $R_{\text{gpu}} = f_{\text{gpu}}/f_{\text{gpu}}^{\text{default}}$ and $R_{\text{mem}} = f_{\text{mem}}/f_{\text{mem}}^{\text{default}}$.
% We use GPU frequency ratio  and memory frequency ratio  to learn frequency-dependent performance scaling while maintaining generalization across different GPU architectures.
Normalizing frequencies allows the model to learn relative scaling trends rather than overfitting to absolute clock values.
The model is trained on the full dataset $\{(U, R_{\text{gpu}}, R_{\text{mem}}, S)\}$ via gradient boosting with squared error loss.
Training across diverse microbenchmarks and frequency configurations enables accurate speedup prediction for unseen kernels based solely on their utilization profiles at the default frequency setting, eliminating the need for exhaustive per-kernel profiling across the entire DVFS space.

\subsection{DVFS Transition Cost Model} \label{sec:dvfs_model}
Accurate characterization of DVFS transition overhead is a prerequisite for effective cost-aware scheduling in DEFT to distinguish between profitable and counterproductive scaling decisions.
%  as the latency can range from hundreds of microseconds to milliseconds and often exceeds the execution time of fine-grained kernels.
Unfortunately, GPU vendors do not publicly disclose DVFS transition mechanisms or timing characteristics.
Prior work~\cite{GPUFreqSwitchLatency2025} has shown that transition latency depends on GPU architecture and voltage regulator design, frequency change magnitude, and transition direction (upscaling vs. downscaling).
As a result, assuming instantaneous transitions or symmetric, linearly scaled costs leads to inaccurate performance estimates and suboptimal scheduling decisions.
The challenge is exacerbated by modern architectures (e.g., Ada Lovelace) that expose fine-grained frequency steps (15 MHz), creating a vast configuration space.
To avoid the prohibitively high cost of exhaustive profiling, we develop a lightweight empirical methodology to construct device-specific transition cost models.

\subsubsection{Measurement Methodology}

Direct measurement of hardware transition latency is challenging because software interfaces such as NVIDIA Management Library (NVML)~\cite{NVML2025} provide only coarse-grained telemetry (e.g., ${\sim}$100 ms sampling resolution on L4 and L40S).
We therefore infer transition latency indirectly by observing the transient impact of frequency scaling on a continuous stream of lightweight kernels. The procedure consists of three stages.
% \textbf{(i) Baseline Characterization:} 
First, for each target frequency $f_{\text{target}}$, we establish a steady-state baseline by executing a sequence of probe kernels and computing the mean execution time $T_{\text{baseline}}(f_{\text{target}})$.
% \textbf{(ii) Asynchronous Transient Probing:} 
Next, we issue a frequency scaling command ($f_{\text{start}} \rightarrow f_{\text{target}}$) and immediately launch a dense stream of asynchronous probe kernels. 
Each probe is timestamped at dispatch and completion using CUDA Events, producing a high-resolution execution trace during the transition.
Finally, we perform post-execution analysis to identify the convergence point. The transition latency is defined as the elapsed time from the first probe launch to the first probe whose execution time stabilizes within a tolerance band ($\pm 5\%$) of $T_{\text{baseline}}(f_{\text{target}})$.

\subsubsection{Model Derivation and Instantiation}

Figure~\ref{fig:dvfs_latency} shows the measured transition latencies on the L4 and L40S GPUs used in our evaluation. 
The data points correspond to a selected set of frequency pairs used to train and validate the multi-tiered transition cost model. 
Each configuration is measured 50 times, and the median value is reported to filter outliers from system jitter.
Measurements across the two types of GPUs reveal several key insights.
First, transition behavior is strongly dependent on GPU design, reflecting differences in voltage regulator design, thermal limits, and power delivery networks~\cite{GPUFreqSwitchLatency2025}. 
Second, transitions exhibit clear directional asymmetry: frequency upscaling is consistently faster than downscaling for the same $\Delta f = |f_{\text{target}} - f_{\text{start}}|$. 
Third, latency does not scale linearly with $\Delta f$; instead, it shows piecewise behavior with discrete thresholds.
Finally, the relationship is non-monotonic: larger $\Delta f$ does not always imply higher latency, reflecting complex interactions across multiple voltage domains and power-management firmware.
Guided by these observations, we construct empirical, GPU-specific transition models by fitting multi-tier piecewise functions, as illustrated in Figure~\ref{fig:dvfs_latency}. 
These models are integrated into the Prediction Engine (\S~\ref{sec:method_overview}) to enable cost-aware evaluation of DVFS decisions during scheduling.

\section{Experimental Setup} \label{sec:setup}

% \subsection{Hardware Platform}
\textbf{Hardware Platform.} The evaluation was conducted on two single-node multi-GPU systems:
(a) a node with eight NVIDIA L4 GPUs, and
(b) a node with four NVIDIA L40S GPUs,
both hosted on dual-socket Intel Xeon Gold 6548N platforms ($2\times64$ cores).
% On L4, the GPU-core frequency can be configured from \SI{210} to \SI{2040}{\mega\hertz} (\SI{15}{\mega\hertz} increment) with memory frequency fixed at \SI{6251}{\mega\hertz}.
% It also supports GPU-core frequencies from 210--\SI{645}{\mega\hertz} when the memory subsystem operates at \SI{405}{\mega\hertz}.
% For L40S, the GPU-core frequency ranges from 210 to \SI{2520}{\mega\hertz} (\SI{15}{\mega\hertz} increment) with memory fixed at \SI{9001}{\mega\hertz}.
% When the memory frequency is reduced to \SI{405}{\mega\hertz}, the GPU-core frequency range narrows to 210--\SI{405}{\mega\hertz}.
Both GPUs expose \SI{15}{\mega\hertz}-granular core-frequency ranges (L4: 210--\SI{2040}{\mega\hertz}, L40S: 210--\SI{2520}{\mega\hertz}) at default memory frequencies (L4: \SI{6251}{\mega\hertz}, L40S: \SI{9001}{\mega\hertz}).
Reducing memory frequency to \SI{405}{\mega\hertz} narrows the core frequency range to 210--\SI{645}{\mega\hertz} on L4 and 210--\SI{405}{\mega\hertz} on L40S.
% Since the \SI{405}{\mega\hertz} memory configuration corresponds to a low-power mode, it is not included in our evaluation.
% The \SI{405}{\mega\hertz} memory state corresponds to a low-power idle mode that cannot be reliably enforced via the application clocks API.
% In practice, the memory remains at the highest frequency throughout execution.
% We therefore exclude this state from scheduling decisions and from our evaluation.
The \SI{405}{\mega\hertz} memory state is a low-power idle mode that cannot be reliably enforced via the application-clocks API (the memory falls back to its maximum frequency at runtime), so we exclude it from scheduling.
% ; both core and memory default to their maximum supported frequencies.
By default, both the GPU core and the memory operate at the maximum supported frequencies.

Both systems run Ubuntu 24.04.3 LTS with NVIDIA driver version 565.57.01.
CUDASTF is compiled using \code{nvcc}~12.4 and \code{g++}~13.3.0.
% All experiments are performed with DVFS control enabled. 
DVFS control is implemented using NVML, specifically through the \code{nvml\-Device\-Set\-Applications\-Clocks()} API call~\cite{NVML2025}.
% Empirical characterization of the L4 and L40S GPUs reveals that the internal hardware power and energy counters update at a granularity of approximately \SI{100}{\milli\second}.
% To ensure measurement fidelity and timely capture of counter increments, we configured the sampling at \SI{50}{\milli\second} intervals via the \code{nvml\-Device\-Get\-Total\-Energy\-Consumption} API.
Empirical characterization shows that the hardware power and energy counters on L4 and L40S GPUs update at ${\sim}$\SI{100}{\milli\second} granularity. 
We therefore sample every \SI{50}{\milli\second} via \code{nvml\-Device\-Get\-Total\-Energy\-Consumption} to capture each increment reliably.

\textbf{Benchmarks.}
We evaluate DEFT using four scientific computing applications with diverse computational characteristics and data access patterns.
\textbf{Cholesky} factorization~\cite{CholeskyTiled2009, abdelfattah2016cholesky} performs blocked dense linear algebra using CUBLAS/CUSOLVER routines ($N{=}32768$, tile size 2048; 816 tasks in total).
Conjugate Gradient \textbf{CG}~\cite{hestenes1952conjugate} solves large sparse linear systems in CSR format using sparse matrix–vector multiplications, dot products, and vector operations (L4: 210M unknowns, 169 tasks in total; L40S: 450M, 175 tasks in total).
Finite-Difference Time-Domain \textbf{FDTD}~\cite{yee1966numerical} simulates electromagnetic wave propagation via 3D stencil updates (L4: $512^3$, L40S: $1024^3$; 200 iterations, 1200 tasks in total).
\textbf{miniWeather}~\cite{norman2020miniweather} models atmospheric dynamics on a 2D domain (L4: $13000{\times}11000$, 360 tasks in total; L40S: $16384{\times}13107$, 240 tasks in total) with heterogeneous domain decomposition, creating irregular workload distribution and asymmetric communication patterns.
% These benchmarks span different domains: linear algebra, electromagnetic simulation, iterative solvers, and computational fluid dynamics, providing comprehensive coverage of common HPC workload patterns.
% \begin{table}[t]
% \centering
% \caption{Benchmark configurations and task graph statistics.}
% \vspace{-0.1cm}
% \label{tab:benchmarks}
% \begin{tabular}{@{}llrc@{}}
% \toprule
% \textbf{Benchmark} & \textbf{Platform} & \textbf{Tasks} & \textbf{Problem Size} \\
% \midrule
% Cholesky & Both & 816 & $N{=}32768$, tile${=}2048$ \\
% \multirow{2}{*}{CG} & L4 & 169 & 210M unknowns \\
%  & L40S & 175 & 450M unknowns \\
% \multirow{2}{*}{FDTD} & L4 & 1200 & $512^3$, 200 iter. \\
%  & L40S & 1200 & $1024^3$, 200 iter. \\
% \multirow{2}{*}{miniWeather} & L4 & 360 & $13000{\times}11000$ \\
%  & L40S & 240 & $16384{\times}13107$ \\
% \bottomrule
% \end{tabular}
% \vspace{-0.2cm}
% \end{table}

\begin{figure*}[!t]
    \centering
    % \vspace{-0.3cm}
    \begin{subfigure}[b]{\textwidth}
        \centering
        \includegraphics[width=\textwidth]{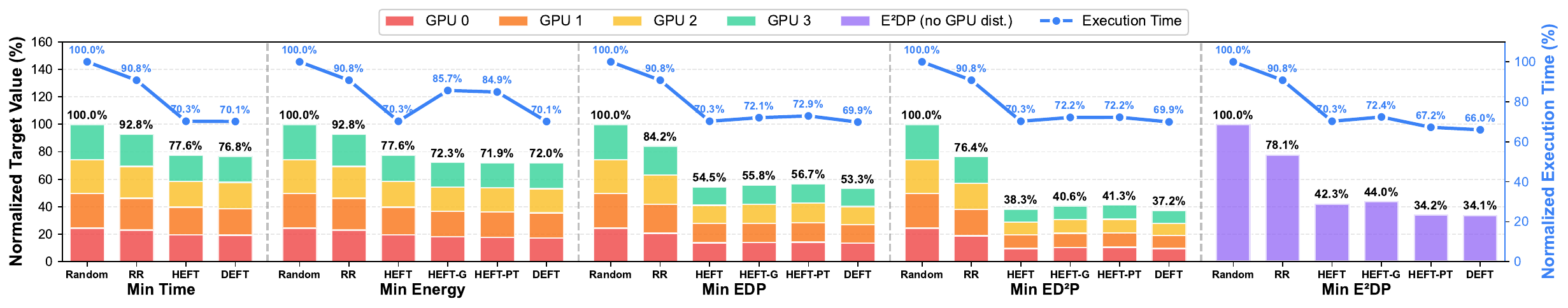}
        \vspace{-0.6cm}
        \caption{Cholesky decomposition}
        \label{fig:L40S_cholesky}
    \end{subfigure}
    
    \begin{subfigure}[b]{\textwidth}
        \centering
        \includegraphics[width=\textwidth]{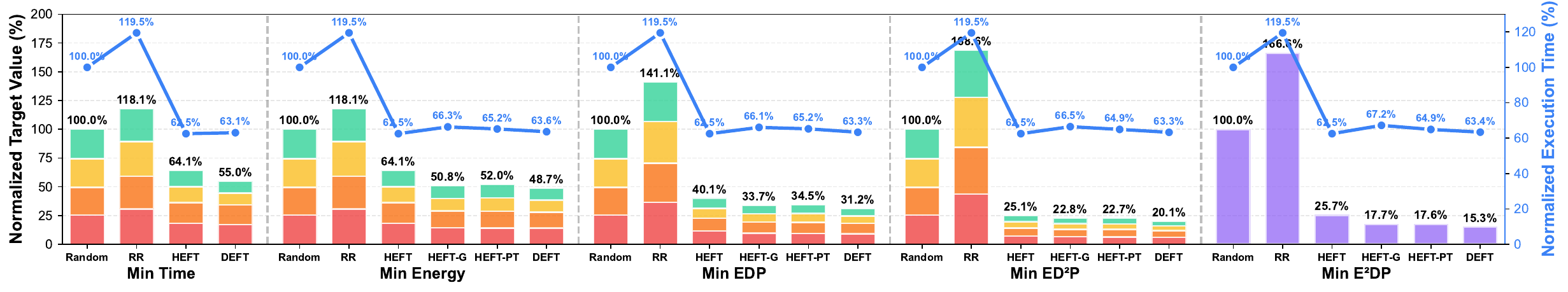}
        \vspace{-0.6cm}
        \caption{Conjugate Gradient}
        \label{fig:L40S_cg}
    \end{subfigure}
    
    \begin{subfigure}[b]{\textwidth}
        \centering
        \includegraphics[width=\textwidth]{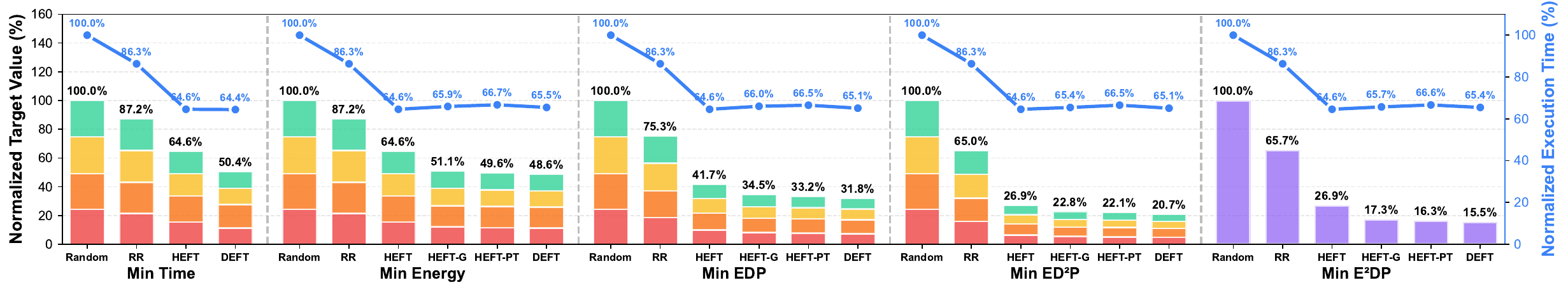}
        \vspace{-0.6cm}
        \caption{Finite-Difference Time-Domain}
        \label{fig:L40S_fdtd}
    \end{subfigure}
    
    \begin{subfigure}[b]{\textwidth}
        \centering
        \includegraphics[width=\textwidth]{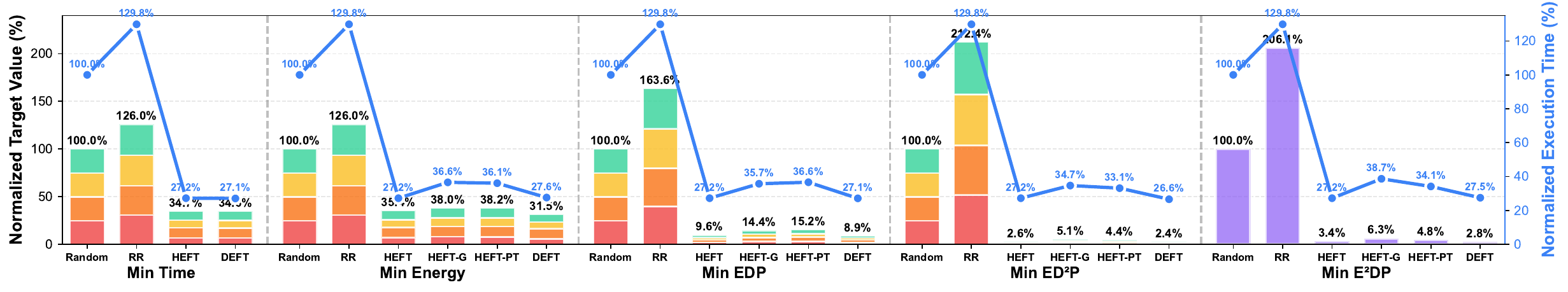}
        \vspace{-0.6cm}
        \caption{miniWeather}
        \label{fig:L40S_miniweather}
    \end{subfigure}
    
    \vspace{-0.3cm}
    \caption{Normalized execution time and energy-performance metrics for the four different schedulers on NVIDIA L40S GPUs (normalized to Random scheduler).
    % Each subplot corresponds to a different optimization target. 
    In the Min Time case, bars show per-GPU energy consumption and the line shows normalized makespan. 
    % bars report per-GPU energy consumption, while the overall makespan is shown as normalized execution time. 
    For energy-related objectives (Min $\mathbf{\mathrm{Energy}}$, $\mathbf{\mathrm{EDP}}$, $\mathbf{\mathrm{E^{2}DP}}$, $\mathbf{\mathrm{ED^{2}P}}$), bars show the normalized objective value. Note that $\mathbf{\mathrm{E^{2}DP}}$ involves squared energy and is computed from total system energy, thus cannot be decomposed per GPU. }
    \label{fig:L40S}
    \vspace{-0.25cm}
\end{figure*}

\section{Evaluation} \label{sec:evaluation}
In this section, we first evaluate how effectively DEFT achieves various energy–performance trade-off objectives across diverse benchmarks and multi-GPU platforms (\S~\ref{sec:eval_effectiveness}).
% assess DEFT's capability to achieve user-specified energy-performance trade-offs under different optimization objectives
We then validate the accuracy of our predictive models (\S~\ref{sec:eval_model_accuracy}).
Finally, we analyze the computational complexity and overhead introduced by the proposed scheduling approach (\S~\ref{sec:eval_overhead}).

\subsection{Effectiveness of DEFT Scheduling}  \label{sec:eval_effectiveness}

We focus the evaluation on multi-GPU scheduling policies because DEFT is designed to jointly optimize task placement and DVFS across devices.
Existing DVFS-based schedulers primarily target single-GPU execution and do not consider task placement, inter-GPU data movement, or multi-device contention, making direct comparison in multi-GPU settings infeasible.
In the single-GPU case, DEFT reduces to a DVFS-aware scheduler; however, the primary contribution lies in coordinating DVFS with multi-GPU task placement under explicit slack and throughput constraints.
% its benefits relative to prior single-GPU DVFS techniques stem from similar predictive modeling and slack-aware decisions, which are orthogonal to the multi-GPU focus of this work.

We evaluate DEFT against three scheduling policies provided by the native CUDASTF runtime: \textbf{Random}, which assigns tasks to devices randomly; \textbf{Round Robin (RR)}, which distributes tasks cyclically; and \textbf{HEFT}, the Heterogeneous Earliest Finish Time algorithm~\cite{Topcuoglu2002HEFT}, which represents a canonical makespan-oriented task graph scheduler and assigns each task to the device that minimizes its earliest finish time. 
None of these policies incorporates DVFS control, relying on the hardware default power management. 
To isolate the contribution of DEFT’s joint scheduling, we implement two DVFS-augmented HEFT variants.
\textbf{HEFT-G} applies a uniform, device-wide frequency that minimizes the target objective across all tasks, representing global frequency tuning without task differentiation.
\textbf{HEFT-PT} applies per-task optimal DVFS after placement, selecting the best frequency for each task independently but without slack or throughput awareness.
Both retain HEFT’s placement decisions and ignore DVFS transition overheads.
We show that DEFT’s gains arise from constrained, globally aware scheduling rather than simply adding DVFS to existing placement strategies.

% To further isolate the contribution of DEFT’s coordinated scheduling strategy, we implement two DVFS-augmented HEFT variants. 
% \textbf{HEFT-G} applies a uniform, device-wide frequency selected to minimize the target objective across all tasks, representing global frequency tuning without task differentiation. 
% \textbf{HEFT-PT} applies per-task optimal DVFS configurations after placement, selecting the best frequency for each kernel independently, but without slack- or throughput-aware constraints. Both variants retain HEFT’s original placement decisions and do not explicitly model DVFS transition overheads.
% Our analysis demonstrate that DEFT’s gains stem from constrained, globally aware scheduling rather than from a simple incremental application of DVFS to existing placement strategies.

% When DVFS is disabled, DEFT’s device selection policy converges to the same task placement as HEFT under fixed frequency. 
% This demonstrates that DEFT preserves the placement quality of established heuristics, while extending them with principled DVFS modeling and multi-objective optimization to improve energy efficiency.

When DVFS is disabled, DEFT’s device selection policy converges to the same task placement as HEFT. 
This has two important implications: (1)~it confirms that DEFT preserves the placement quality of established makespan-oriented heuristics under fixed operating conditions, and (2)~it ensures that the comparison between HEFT and DEFT isolates the contribution of DVFS-aware scheduling.
When DVFS is enabled, DEFT’s scheduling decisions diverge from HEFT because execution times become frequency-dependent, altering device availability and task timing. 
By coordinating placement feasibility and frequency selection, DEFT consistently outperforms HEFT (the fastest baseline) across multiple objectives. 
For example, DEFT reduces energy by 14.8\% on L40S and 4.8\% on L4 and EDP by 9.9\% and 3.7\% (geometric mean), respectively, while remaining within 1.5\% of HEFT's performance.

\subsubsection{Results on NVIDIA L40S GPUs}
% DEFT is configured with five optimization objectives to explore different points in the energy–performance trade-off space.
The evaluation is configured under five optimization objectives: 
% based on the generalized metric $\mathbf{\mathrm{ED^{\beta}P}}$ to explore different points in the energy–performance trade-off space: 
Energy-only ($\beta=0$), $\mathbf{\mathrm{E^2DP}}$ ($\beta=0.5$), EDP ($\beta=1$), $\mathbf{\mathrm{ED^2P}}$ ($\beta=2$), and Time-only ($\beta \to \infty$).
Figure~\ref{fig:L40S} reports the normalized objective values and makespan for the four applications on the four-GPU NVIDIA L40S node.

% In \textbf{Cholesky},
% is a compute-bound workload for which kernel execution at the maximum frequency yields both optimal per-kernel performance and energy on the GPU. 
\textbf{Cholesky.}
Under the time-only objective, DEFT converges to the same decisions as HEFT and achieves comparable performance.
Under energy-related objectives, DEFT reduces total energy by 7.2\% relative to HEFT (2.3\% in EDP, 3.0\% in $\mathbf{\mathrm{ED^{2}P}}$, and 3.9\% in $\mathbf{\mathrm{E^{2}DP}}$) by exploiting task slack for mild frequency reduction, while maintaining similar execution time. 
HEFT-G and HEFT-PT reduce more energy than HEFT by exploiting DVFS; however, they incur execution time inflation due to aggressive frequency scaling without slack or throughput awareness, especially under resource contention. 

% Overall, this demonstrates that modest energy savings are achievable even for compute-dominated workloads without sacrificing performance.

% \textbf{CG} exhibits distinct optimization behavior due to its iterative structure, sparse linear algebra kernels, and limited task-level parallelism. 
\textbf{CG.}
Under the time-only objective, DEFT achieves execution times comparable to HEFT while reducing energy consumption by 14.3\%. 
Under energy-related objectives, DEFT delivers substantial improvements, reducing energy by 24.0\%, EDP by 22.3\%, $\mathbf{\mathrm{ED^{2}P}}$ by 19.7\%, and $\mathbf{\mathrm{E^{2}DP}}$ by 40.6\%, with only a 1.3\% increase in execution time compared to HEFT. 
% These results highlight the effectiveness of slack-aware scheduling: aggressive frequency reduction on non-critical tasks preserves overall progress while yielding significant energy savings. 
Due to limited task-level parallelism in CG, some GPUs periodically become idle.
% DEFT exploits these idle periods by transitioning devices into lower-power states, further reducing energy consumption.
% As the figure shows, the GPUs 2 and 3 consume significantly less energy than GPUs 0 and 1, indicating that DEFT effectively leverages DVFS and idle period management to optimize energy efficiency.
DEFT leverages these intervals to transition idle devices into lower-power states, substantially reducing energy consumption on GPUs 2 and 3, as shown in the per-GPU energy breakdown.
HEFT-PT performs slightly worse than HEFT-G despite per-task frequency selection: frequent DVFS transitions across short sparse kernels introduce additional overhead,
while HEFT-G avoids transition costs but cannot exploit kernel heterogeneity.
% While HEFT-PT is more energy-efficient than HEFT-G by exploiting the diverse task heterogeneity, 
DEFT, by contrast, bounds slowdown using slack and contention awareness, achieving balanced energy savings without degrading performance.

\begin{figure*}[!t]
    \centering
    \vspace{-0.4cm}
    \includegraphics[width=\textwidth]{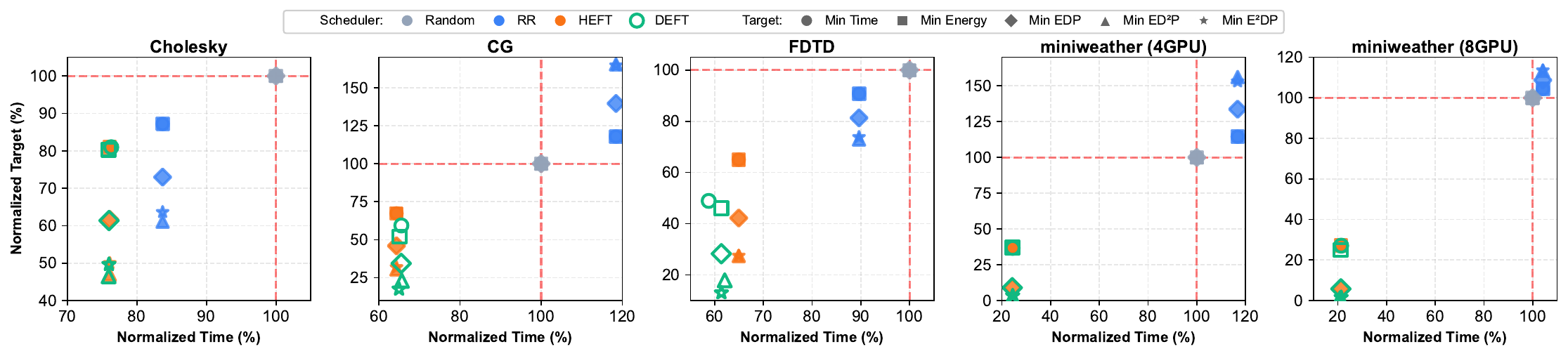}
    \vspace{-0.7cm}
    \caption{Normalized performance and optimization objectives of four scheduling algorithms on NVIDIA L4 GPUs. For Random, Round-robin (RR), and HEFT, the Min Time and Min Energy configurations produce identical data points because both plot energy (y-axis) against makespan (x-axis); these schedulers operate at fixed frequency and thus have the same energy-time coordinates regardless of the stated objective. Consequently, their markers (circles and squares) overlap in the scatter plots.
    % Normalized performance and optimization targets of four different scheduling algorithms on NVIDIA L4 GPUs. For the Random, Round-robin (RR), and HEFT schedulers, the data used for Min Time and Min Energy objectives are identical, therefore, the corresponding markers (circles and squares) overlap in the scatter plots.
    }
    \vspace{-0.25cm}
    \label{fig:L4}
\end{figure*}

% To isolate the benefit of per-task frequency adaptation, we augment HEFT with a uniform oracle frequency set to the average of per-kernel optimal frequencies. This oracle baseline reduces energy by only 5.1\%, versus DEFT’s 24.0\% reduction, 
% % confirming the importance of per-task fine-grained DVFS control.
% confirming that fine-grained, per-task frequency control is essential for exploiting workload-specific scaling behavior.

\textbf{FDTD}.
FDTD is memory-bound and benefits strongly from DVFS. 
When optimizing for performance, DEFT matches HEFT’s execution time while reducing energy consumption by 21.9\%. 
Under energy-related objectives, DEFT achieves 24.7\% reduction in energy, 23.6\% in EDP, 23.0\% in $\mathbf{\mathrm{ED^{2}P}}$, and 42.6\% in $\mathbf{\mathrm{E^{2}DP}}$, with negligible (1.1\%) slowdown in comparison to HEFT.
Although HEFT-G and HEFT-PT improve energy efficiency over HEFT, they incur execution-time penalties. 
HEFT-PT, for instance, reduces energy by 23.1\% but degrades performance by 4\%, reflecting the absence of slack and contention awareness. 
In contrast, DEFT maintains near-optimal performance while achieving additional energy savings of approximately 2–5\% over both HEFT variants.
% These results confirm that DVFS-aware scheduling can significantly improve energy efficiency for memory-bound workloads without compromising performance, and that DEFT effectively exploits this opportunity at task granularity.

\textbf{miniWeather} features irregular workload distribution due to heterogeneous domain decomposition.
% represents an atmospheric dynamics simulation with irregular workload distribution across the two-dimensional computational domain. 
% To stress-test DEFT’s ability to handle load imbalance, we configure the simulation with heterogeneous domain decomposition, resulting in asymmetric task execution times and non-uniform inter-task communication patterns. 
Random and RR perform poorly because device-agnostic placement induces excessive inter-GPU communication. 
HEFT significantly improves performance by prioritizing data locality. 
When optimizing for performance, DEFT matches HEFT by selecting devices that prioritize earliest start time and locality.
Its joint optimization of placement and frequency further reduces energy by 11.1\% with only a 1.4\% performance overhead. 
% However, DEFT's joint optimization of placement and frequency selection enables additional energy savings: compared to HEFT, DEFT reduces energy consumption by 10.9\% while maintaining comparable performance, with only 1.4\% performance overhead.
Under energy-related objectives, DEFT delivers 8.0\% reduction in EDP, 9.5\% in $\mathbf{\mathrm{ED^{2}P}}$, and 18.0\% in $\mathbf{\mathrm{E^{2}DP}}$.
The irregular workload creates substantial slack, which DEFT selectively exploits by lowering frequency on non-critical tasks while preserving the critical path.
In contrast, HEFT-G and HEFT-PT slow both critical and non-critical tasks, thereby significantly extending execution time.
This inflated makespan increases system idle energy and reduces overall efficiency. 
For example, compared to HEFT-G, DEFT improves performance by 17\% and reduces energy by 25\%.
% demonstrating the importance of slack-bounded and contention-aware frequency control.
% The irregular workload distribution creates substantial task slack opportunities, while some GPUs execute critical-path tasks at high frequency, DEFT exploits available slack on non-critical tasks to reduce frequency on other devices.
% This demonstrates DEFT's effectiveness in exploiting workload irregularity: the scheduling framework automatically identifies and leverages slack created by load imbalance, converting what would otherwise be wasted opportunity into tangible energy savings without extending the critical path.

% Overall, these results show that conventional scheduling policies leave substantial energy optimization opportunities unexploited. 
% By jointly optimizing task placement and frequency selection using slack-aware, cost-aware analysis and idle period management, DEFT consistently achieves superior energy efficiency while preserving performance across diverse workloads.
We also evaluated a Mixture-of-Experts workload; however, sustained high utilization triggers hardware power limiting and GPU Boost, which overrides software DVFS control, restricting DEFT’s ability to adjust frequency. 
This behavior identifies compute-saturated, power-bound workloads as a boundary condition for DVFS-based optimization.

\subsubsection{Results on NVIDIA L4 GPUs}
Figure~\ref{fig:L4} summarizes the results for all four benchmarks on the 8-GPU L4 node using scatter plots in normalized target–time space, where points closer to the bottom-left corner indicate better overall trade-offs. 
Across all benchmarks, Random and Round Robin schedulers consistently occupy upper right regions of the plot, reflecting their inability to account for task characteristics, device heterogeneity, or energy--performance trade-offs. 
HEFT achieves strong performance-oriented results in most of the benchmarks.
In contrast, DEFT consistently achieves improved trade-offs across multiple optimization targets. 

For compute-bound workloads such as \textbf{Cholesky}, DEFT slightly outperforms HEFT (by ${\sim}$1\%) under energy-related objectives.
For \textbf{CG}, DEFT significantly improves energy-related metrics (e.g.,~22.7\% energy reduction with only 1.2\% impact on execution time), demonstrating the effectiveness of slack-aware DVFS decisions and idle power management on iterative and memory-sensitive workloads.

Notably, for \textbf{FDTD}, DEFT not only improves energy efficiency but also outperforms HEFT when optimizing for execution time, achieving a shorter makespan. 
This behavior can be attributed to the interaction between workload characteristics and GPU power management dynamics.
FDTD is memory-bound and exhibits relatively low and fluctuating compute utilization.
As a result, operating persistently at the maximum GPU frequency does not yield optimal performance.
Without explicit DVFS control, the GPU's hardware power and thermal management mechanisms autonomously adjust the operating frequency in response to instantaneous load variations, leading to pronounced frequency oscillations (e.g., 531 frequency transitions observed under HEFT vs. 88 under DEFT) and execution-time variability.
In contrast, DEFT proactively selects a slightly lower but stable operating frequency based on predictive modeling, suppressing runtime frequency fluctuations and improving execution stability.
This highlights an advantage of proactive DVFS management: it uncovers performance opportunities inaccessible when frequency control is left to hardware defaults.

% Under HEFT, which lacks DVFS control on L4 GPUs, hardware power and thermal management mechanisms frequently adjust the operating frequency in response to instantaneous load variations, leading to pronounced frequency oscillations (e.g., 531 frequency transitions observed under HEFT versus 88 under DEFT) and execution-time variability.
% This highlights an important advantage of DEFT: it uncovers performance opportunities that are inaccessible to time-only schedulers such as HEFT.
% through coordinated placement and frequency control.

We evaluate \textbf{miniWeather} (average task parallelism of 6) across 4-GPU and 8-GPU configurations to examine both resource-con\-strained and resource-abundant scenarios, as well as the scalability of DEFT under varying hardware availability.
With 4 GPUs, the system operates under resource contention, and DEFT tightens slack bounds to limit frequency scaling and prevent excessive execution-time inflation. 
As a result, DEFT achieves performance and energy efficiency comparable to HEFT. 
Disabling throughput awareness degrades performance by 9.6\% relative to HEFT, confirming that contention-aware slack refinement is essential under resource pressure.
With 8 GPUs, the system has sufficient resources to accommodate the workload, enabling DEFT to exploit task slack more effectively.
As a result, DEFT achieves 8.4\% energy reduction compared to HEFT while maintaining comparable performance.

% Overall, these results demonstrate that DEFT consistently dominates conventional scheduling policies on the L4 platform, delivering superior energy–performance trade-offs while, in some cases, improving performance beyond state-of-the-art time-oriented schedulers.

% The results validate three key claims:
% \begin{enumerate}
%     \item \textbf{Configurability:} DEFT's parameterized objective function ($E^\alpha D^\beta P$) successfully enables users to navigate the energy-performance trade-off space according to application requirements.
%     \item \textbf{Dominance:} DEFT consistently outperforms the naive schedulers, demonstrating that joint optimization of task placement and frequency selection is essential for achieving optimal energy-performance trade-offs.
%     \item \textbf{Slack Exploitation:} The substantial gap between DEFT and HEFT in the performance dimension confirms that slack-aware frequency scaling avoids extending the critical path, enabling energy savings without proportional performance degradation.
% \end{enumerate}

\subsection{Model Accuracy Validation} \label{sec:eval_model_accuracy}

We validate the proposed power and performance models by comparing the predictions against measured values across the full spectrum of DVFS configurations over the four applications. 
We use Mean Absolute Percentage Error (MAPE) as the evaluation metric, defined as $\mathrm{MAPE} = \frac{1}{N} \sum_{i=1}^{N} \left| \frac{y_i - \hat{y}_i}{y_i} \right| \times 100\%$,
% \[
% \mathrm{MAPE} = \frac{1}{N} \sum_{i=1}^{N} \left| \frac{y_i - \hat{y}_i}{y_i} \right| \times 100\% ,
% \]
where \( y_i \) and \( \hat{y}_i \) denote the measured and predicted values.
Table~\ref{tab:model_mape} reports the per-application MAPE of both models, averaged across the L4 and L40S platforms.
The analytical power model maintains an MAPE of 8.1\%, while the performance model achieves an MAPE of 4.3\% across all benchmarks and GPU platforms, indicating that the proposed models capture the non-linear scaling characteristics of the hardware and provide sufficiently accurate predictions to support effective energy-aware scheduling decisions.

\begin{table}[t]
\centering
\caption{Per-application MAPE (\%) of the power and performance models, averaged across L4 and L40S.}
\vspace{-0.2cm}
\label{tab:model_mape}
\begin{tabular*}{\columnwidth}{@{\extracolsep{\fill}}lcccc@{}}
\toprule
\textbf{Metric} & \textbf{Cholesky} & \textbf{CG} & \textbf{FDTD} & \textbf{miniWeather} \\
\midrule
Power       & 6.7 & 8.3 & 9.5 &  7.9   \\
Performance & 3.8 & 4.2 & 4.7 &  4.5   \\
\bottomrule
\end{tabular*}
\vspace{-0.2cm}
\end{table}

% ### 6. __Model Accuracy Validation__ (Section 4 - Models)

% __Purpose:__ Validate predictive models __Content:__

% - Scatter plot: Predicted vs Actual (Power)
% - Scatter plot: Predicted vs Actual (Performance)
% - Show R², MAPE
% - Different colors for different GPU models

% __Why:__ Builds confidence in model-driven approach

% ---

% ### 8. __Overhead Breakdown__ (Evaluation)

% __Purpose:__ Show scheduling overhead is <1% __Content:__

% - Stacked bar chart: [Scheduling | Execution | Transfers | DVFS]
% - Different workloads

% __Why:__ Addresses potential reviewer concern about overhead

\subsection{Scheduling Overhead Analysis} \label{sec:eval_overhead}

A critical concern for any online scheduling system is the runtime overhead introduced by scheduling decisions.
We analyze DEFT's overhead from both theoretical complexity and empirical measurement perspectives.

% \subsubsection{Computational Complexity}
As described in \S~\ref{sec:scheduling_algo}, DEFT employs a two-phase approach.
%  that reduces the scheduling complexity from $O(|\mathcal{D}| \times |\mathcal{F}|)$ to $O(|\mathcal{D}| + |\mathcal{F}|)$ per task.
Phase 1 iterates over $|\mathcal{D}|$ devices to select the target device,
%  based on temporal availability.
phase 2 evaluates $|\mathcal{F}|$ frequency configurations for the selected device.
%  to minimize the user-specified objective function.
For a task graph with $N$ tasks, the overall complexity is $O(N \times (|\mathcal{D}| + |\mathcal{F}|))$.
In our experimental platforms, $|\mathcal{D}| \in \{4, 8\}$ and $|\mathcal{F}|$ ranges from 122 to 154 discrete DVFS configurations.
% , both small constants that yield effectively linear scaling with task count.
Since both are determined by the hardware platform and independent of the application, scheduling cost scales linearly with task count for a given system configuration.

% \subsubsection{Scheduling Overhead}
We evaluate the runtime overhead of DEFT by measuring the average scheduling time spent per task, rather than reporting a percentage of total execution time.
% Percentage-based metrics are highly sensitive to task granularity and can be misleading in task-based applications, where fine-grained workloads naturally amplify relative scheduling costs.
Percentage-based metrics are sensitive to task granularity and can be misleading, where fine-grained workloads amplify relative scheduling costs.
% Reporting per-task overhead isolates the scheduler's computational cost from application-specific task characteristics, providing a fair assessment across workloads with varying task granularities.
On the L4 and L40S, DEFT incurs an average scheduling overhead of 61~$\mu$s and 89~$\mu$s per task, respectively.
It is 2--3 orders of magnitude smaller than typical kernel execution times in our benchmarks (7--112\,ms), confirming DEFT's viability for online scheduling.
This overhead is incurred only at task submission time and is independent of task execution time, scaling primarily with $|\mathcal{D}|$ and $|\mathcal{F}|$.
% DEFT is designed with a low-complexity heuristic that operates at a microsecond scale. Consequently, 

% ensuring its viability for online scheduling across a broad range of multi-GPU workloads.

% We instrument DEFT to measure the wall-clock time spent in scheduling decisions for each task, including:
% (1) device selection (Phase 1): computing $T_{\text{earliest}}^d$ and $\Delta E_{\text{transfer}}^d$ for all devices;
% (2) frequency optimization (Phase 2): evaluating the cost function across all DVFS configurations for the selected device;
% (3) model inference: querying the performance and power predictors for execution time and power estimates; and
% (4) state updates: updating device availability, MSI coherence states, and frequency configurations.

% Table~\ref{tab:scheduling_overhead} summarizes the average per-task scheduling overhead across all benchmarks.
% On the L40S platform (4 GPUs, 154 DVFS configurations), DEFT incurs [X] $\mu$s per task on average.
% On the L4 platform (8 GPUs, 122 DVFS configurations), the overhead increases to [Y] $\mu$s per task, reflecting the larger device search space.
% These measurements confirm that DEFT's scheduling overhead scales primarily with the number of devices and frequency configurations, not with task execution time.

\section{Related Work} \label{sec:relatedwork}
Our work relates to two primary research areas: GPU power and performance modeling and energy-aware runtime systems.
We structure the discussion around these topics and highlight how our approach differs from prior work.
% Our work relates to GPU power and performance modeling and energy-aware runtime systems, and we structure the discussion around these topics.

\textbf{GPU Power and Performance Modeling.}
Several prior works have developed power and performance models to characterize the impact of DVFS on GPU energy consumption and execution time.
Empirical studies show that the optimal GPU frequency depends strongly on application characteristics and architecture~\cite{GPUDVFSSurvey2017}.
Analytical and machine-learning models predict power and performance across the DVFS space from hardware counters collected at a single frequency, using functional-component decomposition~\cite{GPUPowerModel2018, GPUPowerDVFS2019}, neural networks with kernel clustering~\cite{GPUPowerPerfModelML2015}, microbenchmark-calibrated architectural parameters~\cite{GPUPerfModel2020}, or offline multi-objective frequency selection~\cite{GPUFreqSelection2022}.
To avoid runtime profiling, static approaches derive features directly from kernel code, including OpenCL operations~\cite{OpenCLEnergyFreqScaling2019}, LSTM-encoded PTX sequences~\cite{PowerModelPTX2019}, and DCGM--PTX fusion for per-application frequency selection~\cite{DSO_IWQoS2024}.

% Our work extends beyond modeling to integrate prediction models into a runtime scheduling framework that jointly optimizes energy and performance in multi-GPU environments.
% While static analysis eliminates profiling overhead, our approach uses lightweight one-time runtime profiling to capture actual hardware utilization patterns, which proves more accurate for diverse workloads and enables adaptation to runtime system conditions that static analysis cannot capture.

% ADDED: Discuss DVFS transition modeling gap
% \textbf{DVFS Transition Overhead.}
% Most prior DVFS optimization work either ignores frequency transition overhead or assumes simplified models (e.g., constant latency or linear scaling).
% Recent empirical studies~\cite{GPUFreqSwitchLatency2025} have demonstrated that GPU DVFS transitions exhibit asymmetric, non-linear, and architecture-dependent characteristics, with latencies ranging from hundreds of microseconds to milliseconds.
% % ADDED: Our contribution
% Our work explicitly addresses this gap by developing empirical piecewise models that capture the complex transition behavior (directional asymmetry, discrete thresholds, non-monotonicity) and integrate these costs directly into scheduling decisions, ensuring overhead-aware optimization.

\textbf{Energy-Aware Runtime Systems.}
Several works explore integrating DVFS control into GPU programming models and runtime systems to enable energy-aware execution.
SYnergy~\cite{SYnergySC2023} extends SYCL with a per-kernel energy interface that drives frequency selection through compile-time analytical models, and phase-based scaling~\cite{PhaseBasedFS2025} groups similar kernels under a uniform frequency to amortize DVFS overhead.
Complementary efforts target placement rather than DVFS: unbalanced power capping with StarPU~\cite{PowerCapEnergy2025, StarPU}, genetic-augmented HEFT to mitigate kernel interference~\cite{MultiObjGPUKernSched2024}, SM-level load balancing~\cite{HuangGPUBalance2020}, and core/warp throttling for memory-intensive workloads~\cite{song2014energy}.
For heterogeneous multi-CPU systems, JOSS~\cite{JOSS2023} and its follow-up SWEEP~\cite{SWEEP2024} jointly schedule tasks with CPU/memory DVFS under energy--performance constraints.

% In summary, existing work addresses GPU modeling and DVFS-based energy-aware scheduling, yet these techniques are largely confined to single-GPU settings or treat task placement and frequency control as separate concerns.
% DEFT fills this gap by integrating task placement and DVFS decisions within a unified, cost-aware runtime framework for multi-GPU task graphs.
In contrast to prior work, which confines DVFS and placement to single-GPU settings or treats them as separate concerns, DEFT integrates both within a unified, cost-aware runtime framework for multi-GPU task graphs.

\section{Conclusion} \label{sec:conclusion}
% Multi-GPU architectures, while delivering unprecedented computational throughput, present challenges for energy-efficient execution due to the tightly coupled effects of task placement, DVFS configuration, and inter-GPU data movement.
% In this work, we presented DEFT, an energy-aware scheduling framework that jointly coordinates device selection and per-GPU DVFS configuration for task-based multi-GPU applications to achieve user-specified energy–performance trade-offs with minimal performance degradation.
% DEFT integrates slack-awareness, throughput-awareness, and explicit models of task execution cost, inter-GPU data transfer, and DVFS transition overheads within a unified cost model, enabling coordinated, task-granularity–aware decisions that preserve task parallelism, data locality while preventing excessive slowdown under contention.
% Experimental results on NVIDIA L4 and L40S multi-GPU nodes across diverse applications demonstrate substantial improvements across multiple energy–performance objectives compared to native CUDASTF scheduling policies, validating the effectiveness of DEFT in unlocking significant energy efficiency gains in multi-GPU runtimes while maintaining high system throughput.

Multi-GPU architectures deliver massive computational throughput but pose significant challenges for energy-efficient execution, due to the tight coupling between task placement, DVFS configuration, and inter-GPU data movement.
This work presents DEFT, an energy-aware scheduling framework that jointly coordinates task-to-device placement and per-GPU DVFS configuration to achieve user-specified energy--performance trade-offs in task-based multi-GPU applications.
By integrating slack and throughput awareness with explicit modeling of execution cost, data movement and DVFS transition cost within a unified cost model, DEFT enables coordinated, task-granularity–aware decisions that preserve task parallelism and data locality while avoiding excessive slowdown under contention.
% It is guided by prediction models and integrates slack and throughput awareness with explicit modeling of execution cost, data movement and DVFS transition cost within a unified cost model, DEFT effectively identifies the opportunity of task slack, idleness and task heneterogenety for improving energy efficiency while preventing performance slowdown. 
% By incorporating slack and throughput awareness with explicit modeling of execution cost, data movement and DVFS transition cost, DEFT effectively coordinates placement and frequency at task granularity without compromising performance. 
% DEFT integrates three predictive models to support runtime scheduling decisions. 
The evaluation on NVIDIA L4 and L40S multi-GPU systems demonstrates that DEFT consistently improves energy efficiency across multiple objectives compared to native CUDASTF scheduling policies, while maintaining high performance.

\begin{acks}
This work was supported in part by the Swedish Foundation for Strategic Research (SSF) through PRIDE (CHI19-0048), the Swedish Research Council (VR) through P4PIM (2020-04892), and the EuroHPC Joint Undertaking through DARE under grant agreement No. 101202459. Computational resources were provided by the Minerva cluster at the Department of Computer Science and Engineering, Chalmers University of Technology.
\end{acks}

\bibliographystyle{ACM-Reference-Format}
\bibliography{sample-base.bib}

\end{document}